\documentclass[acmtog, preprint]{acmart}
\acmSubmissionID{247}
\usepackage{booktabs} 

\setcitestyle{square}

\usepackage{wrapfig}
\usepackage{float}
\usepackage{subcaption}
\usepackage{color}
\usepackage[utf8x]{inputenc}
\usepackage[colorinlistoftodos]{todonotes}
\usepackage{xargs}                  
\usepackage{algpseudocode,algorithmicx}
\usepackage[normalem]{ulem}

\usepackage[ruled]{algorithm2e} 

\SetAlFnt{\small}
\SetAlCapFnt{\small}
\SetAlCapNameFnt{\small}
\SetAlCapHSkip{0pt}
\IncMargin{-\parindent}

\definecolor{gray}{rgb}{0.5,0.5,0.5}
\definecolor{green}{rgb}{0, 0.6, 0}
\definecolor{orange}{rgb}{1, 0.5, 0}
\definecolor{mahogany}{rgb}{0.75, 0.25, 0.0}
\definecolor{purple}{rgb}{0.6, 0, 0.6}
\definecolor{darkgreen}{rgb}{0, 0.4, 0}
\definecolor{dijon}{rgb}{0.8, 0.5, 0}
\definecolor{ballblue}{rgb}{0.1, 0.6, 0.8}

\newcommand{\com}[1]{}

\newcommandx{\unsure}[2][1=]{\todo[linecolor=red,backgroundcolor=red!25,bordercolor=red,#1]{#2}}
\newcommandx{\change}[2][1=]{\todo[linecolor=blue,backgroundcolor=blue!25,bordercolor=blue,#1]{#2}}
\newcommandx{\info}[2][1=]{\todo[linecolor=OliveGreen,backgroundcolor=OliveGreen!25,bordercolor=OliveGreen,#1]{#2}}
\newcommandx{\improvement}[2][1=]{\todo[linecolor=Plum,backgroundcolor=Plum!25,bordercolor=Plum,#1]{#2}}
\newcommandx{\thiswillnotshow}[2][1=]{\todo[disable,#1]{#2}}

\newcommand{\old}[1]{}
\newcommand{\revised}[1]{{#1}}

\acmJournal{TOG}

\begin{teaserfigure}  
 \includegraphics[width=\textwidth]{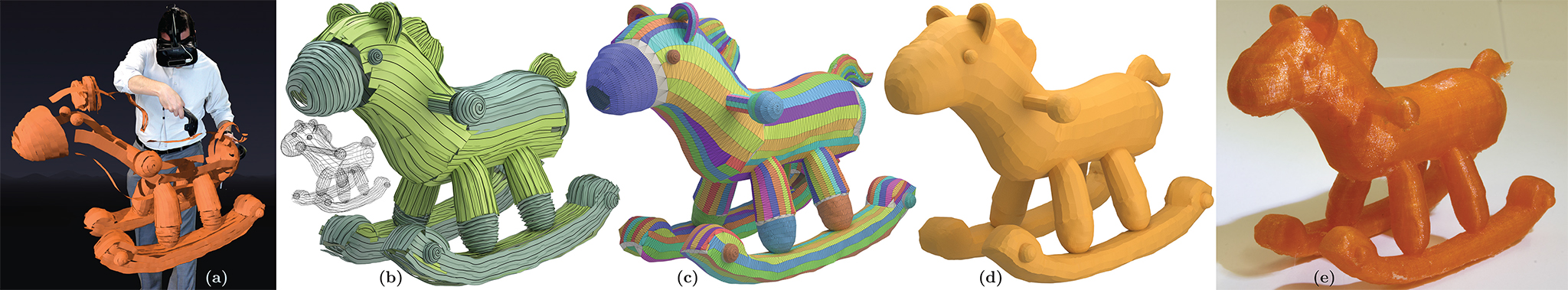}
   \caption{(a) Drawing 3D strokes using a VR brush. (b) Completed 3D brush-stroke drawing with central stroke polylines drawn in black and ribbon color reflecting normal orientation (green for front, turquoise for back, inset shows poylines alone). (c) triangle mesh strips connecting adjacent stroke polylines (multicolor), and  gray triangle strips that complete the surface connecting differently directed stroke groups. (d) Final output. (e) Fabricated model. Input drawing: $\copyright$ Jafet Rodriguez.}
\label{fig:teaser}
\end{teaserfigure}

\begin{document}
\setcopyright{acmlicensed}
\acmJournal{TOG}
\acmYear{0}\acmVolume{0}\acmNumber{0}\acmArticle{0}\acmMonth{0}
\acmDOI{10.1145/nnnnnnn.nnnnnnn}

\title{SurfaceBrush:  From Virtual Reality Drawings to Manifold Surfaces}

\author{Enrique Rosales}
\affiliation{%
  \institution{University of British Columbia}
  \country{Canada}
}
\email{albertr@cs.ubc.ca}
\affiliation{%
  \institution{Universidad Panamericana}
  \department{Facultad de Ingenier\'{i}a}
  \city{Zapopan}
  \state{Jalisco}
  \postcode{45010}
  \country{M\'{e}xico}
}

\author{Jafet Rodriguez}
\affiliation{%
  \institution{Universidad Panamericana}
  \department{Facultad de Ingenier\'{i}a}
  \city{Zapopan}
  \state{Jalisco}
  \postcode{45010}
  \country{M\'{e}xico}
  }
\email{arodrig@up.edu.mx}

\author{Alla Sheffer}
\affiliation{%
  \institution{University of British Columbia}
  \country{Canada}
  }
\email{sheffa@cs.ubc.ca}

\begin{abstract} 
Popular Virtual Reality (VR) tools allow users to draw varying-width, ribbon-like 3D brush \old{stokes} \revised{strokes} by moving a hand-held controller in 3D space.  Artists frequently use dense collections of such strokes to draw virtual 3D shapes. We propose {\em SurfaceBrush}, a surfacing method that converts such VR drawings into \old{user intended} \revised{user-intended} manifold free-form 3D surfaces, providing a novel approach for modeling 3D shapes. The inputs to our method consist of dense collections of artist-drawn stroke ribbons described by the positions and normals of their central polylines, and ribbon widths. These inputs are highly distinct from those handled by existing surfacing frameworks and exhibit different sparsity and error patterns, necessitating a novel surfacing approach. We surface the input stroke drawings by identifying and leveraging local coherence between nearby artist strokes.  In particular, we observe that strokes intended to be adjacent on the artist imagined surface often have similar tangent directions along their respective polylines. We leverage this local stroke direction consistency  by casting the computation of the user-intended manifold surface as a constrained matching problem on stroke polyline vertices and edges. We first detect and smoothly connect adjacent similarly-directed sequences of stroke edges producing one or more manifold partial surfaces. We then complete the surfacing process by identifying and connecting adjacent similarly directed edges along the borders of these partial surfaces. 
We confirm the usability of the SurfaceBrush interface and the validity of our drawing analysis via an observational study. We validate our stroke surfacing algorithm by demonstrating an array of manifold surfaces computed by our framework starting from a range of inputs of varying complexity, and  by comparing our outputs to reconstructions computed using alternative means. 
\vspace{-6pt}
\end{abstract}

\keywords{Virtual Reality, 3D drawing, surface modeling, surface reconstruction\vspace{-6pt} }

\maketitle

\section{Introduction}
\label{sec:intro}

Humans frequently communicate 3D shapes via 2D sketches  or drawings, inspiring the development of modeling interfaces that employ such drawings as inputs~\cite{OSSJ09}.
Virtual Reality (VR) systems support real-time capture and visualization of human 3D gestures enabling users to draw surfaces directly in 3D space (Figure~\ref{fig:teaser}a). Using such drawings as input for 3D modeling can sidestep the main algorithmic challenge of \old{the} 2D sketch-based modeling methods -- the need to derive 3D information from a 2D input. Effectively leveraging the opportunity provided by VR interfaces requires  modeling frameworks capable of processing the types of 3D drawings users can comfortably provide using these tools. Our observations show that artists using the VR medium frequently  depict complex  free-form 3D geometries using collections of dense, ruled surface  {\em brush strokes}  traced in 3D space (Figure~\ref{fig:teaser}b) \cite{TiltBrushRepo}. Our {\em SurfaceBrush} framework algorithmically converts VR brush stroke drawings into manifold surface meshes describing the user-intended geometry (Figure~\ref{fig:teaser}d), enabling downstream applications such as 3D printing \old{(}(Figure~\ref{fig:teaser}e).

Users of VR systems, such as TiltBrush~\shortcite{TiltBrush} or Drawing on Air~\cite{Keefe:2007}, trace strokes using a handheld controller. These systems then automatically translate controller locations into polyline stroke vertex positions and controller orientations into stroke normals. They subsequently render the captured input as virtual ribbons, or ruled surface strips, of user-specified width (Figure~\ref{fig:teaser}a). The rendered ribbons are centered around the captured stroke polyline positions and their orientation reflects the captured stroke normals. Our experiments show that both artists and non-experts can easily, quickly, and effectively communicate their envisioned 3D surfaces using this interface by drawing dense brush strokes that cover the surface of the intended shapes (Section~\ref{sec:user_study}).

\begin{figure}
\includegraphics[width=\columnwidth]{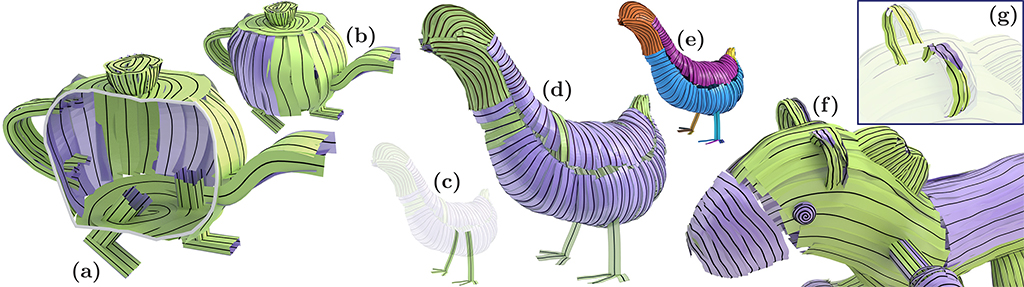}
\caption{Brush stroke \old{drawing} \revised{drawings} are characterized by 
strokes with partially overlapping ribbons, and locally similar stroke tangent directions (b,d,f). These directions change abruptly between different surface regions (e); stroke normal orientations are often inconsistent  (front facing ribbons rendered in green, back facing in purple) (b,d,f), inexact and sometimes erroneous (g); stroke groups frequently intersect ``inside'' the model (see cutout view) (a); and drawings occasionally contain isolated strokes (c). Teapot and horse: $\copyright$ Jafet Rodriguez, chicken: $\copyright$ Elinor Palomares.}  
\label{fig:challenges}
\end{figure}

\begin{figure*}
 \includegraphics[width=\textwidth]{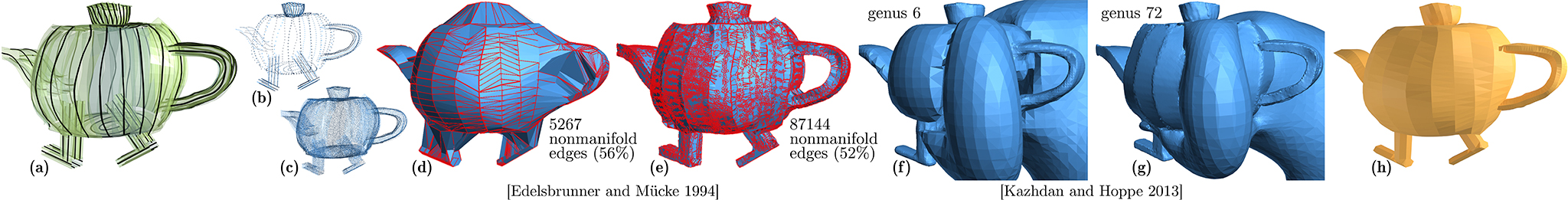}
\caption{Treating strokes' (a) polyline vertices  as unorganized points with normals (b) and using those as input to state of the art reconstruction from point-clouds methods (d,f) produces inadequate results with multiple artifacts, such as high percentages of non-manifold edges \cite{Edels} (d), unnecessarily high genus, and arbitrary deviation from the input \cite{Kazhdan:2013} (f). Densely sampling points along the ruled ribbons (c) and using these samples plus normals as reconstruction input produces surfaces which exhibit similar artifacts (e,g). Our output (h) accurately captures user's intent. Input drawing: $\copyright$ Jafet Rodriguez.}

\label{fig:compare}
\end{figure*}

Adopting this interface for surface modeling necessitates algorithmically reconstructing the user-intended 3D surfaces from the dense set of brush strokes drawn by the users. Each stroke is defined by the vertex positions and normals along its central polyline and has an \old{associate} \revised{associated} width. 
This input format is distinctly different from those processed by existing surface reconstruction methodologies, and exhibits different error and sparsity patterns (Section~\ref{sec:input}). In particular, artist drawings (see e.g. Figure~\ref{fig:challenges}) have inconsistent stroke normal orientations and partially overlapping strokes; they frequently contain intersecting stroke groups and may \revised{exhibit} \old{contain} \old{isloated} \revised{isolated} outlier strokes. Due to these artifacts, \old{exisisting} \revised{existing} surfacing methods are inadequate for our needs (Section~\ref{sec:related}).
In particular, using polyline vertices or densely sampled points on the ribbons as input to methods for reconstruction from point clouds fails to produce the desired surfaces (Figure~\ref{fig:compare}).

\setlength{\belowcaptionskip}{-20pt}
\setlength{\belowcaptionskip}{0pt}

SurfaceBrush reconstructs an intended surface from the input brush strokes by interpolating sequences of edges along the stroke polylines. It determines the edges to include and the connectivity between them by leveraging local consistency between the drawn strokes (Section~\ref{sec:data}).
The key observation it utilizes is that, when depicting 3D shapes using a VR brush,  users typically adopt a strategy that resembles the action sequence commonly used when applying top paint to 3D objects using a paint brush. Specifically, users often draw contiguous patches on the target surface using side-by-side strokes with similar tangent directions and change stroke directions when switching between different parts of the drawing (Figure~\ref{fig:challenges}).  
These \old{observation} \revised{observations} argue for a surfacing strategy that prioritizes connections between side-by-side strokes with similar tangents. Following this argument, SurfaceBrush computes the output surface using a \old{two step} \revised{two-step} process (Section~\ref{sec:overview}). First, it forms {\em inter-stroke} mesh strips, by detecting and connecting side-by-side stroke sections, or sequences of edges (Figure~\ref{fig:teaser}c, multicolor, Section~\ref{sec:local}). Then, it closes the gaps between the partial surfaces consisting of a union of such strips by connecting adjacent sections along their boundaries (Figure~\ref{fig:teaser}c, gray, Section~\ref{sec:global}).
The core challenge in employing this strip-based surfacing approach is to identify, or to match, the best stroke sections to connect
in the first stage of the process and the best boundary sections to connect in the second. This challenge is augmented by our goal of producing {\em manifold} output surfaces while overcoming artifacts present in the data (Figure~\ref{fig:challenges}).

We formulate both matching problems using a discrete constrained optimization framework (Section~\ref{sec:local}). We efficiently solve them by  
first relaxing the manifoldness constraints, obtaining locally optimal \revised{(}but not necessarily globally compatible\revised{)} vertex-to-vertex matches (Sections~\ref{sec:asses},~\ref{sec:matching}). 
We use the obtained vertex-to-vertex matches to identify corresponding stroke sections and to connect these sections using triangle strips (Sections~\ref{sec:mesh}).
We eliminate non-manifold artifacts in this mesh  using a correlation clustering framework that determines which triangles should remain in the mesh and which should be removed (Section~\ref{sec:bijective}).
This  process robustly reconstructs user-intended, manifold surfaces from complex drawings, such as the horse (Figure~\ref{fig:teaser}, 298 strokes, 20K vertices) in under a  minute.

We validate the SurfaceBrush modeling framework by evaluating both our choice of inputs and the method we propose for processing those. 
We conduct a user study which confirms that experts and non-experts alike can effectively use brush strokes to visually communicate  free-form surfaces in a VR environment,
and validates our observation about users preference for depicting surfaces using a set of patches drawn using similarly directed strokes (Section~\ref{sec:study}). We confirm the robustness of the SurfaceBrush surfacing algorithm by demonstrating a range of reconstructed surfaces created from inputs of different complexity produced by artist and amateur users and compare the results to those produced by state of the art alternatives (Section~\ref{sec:results}). These experiments confirm that, while our outputs are consistent with the artist-intended surface geometry, the results of alternative methods are not. 

Our overall contribution is a new VR drawing-based modeling framework that allows experts and amateurs alike to create complex free-form 3D models via an easy-to-use interface.  The technical core of our system is a new surfacing algorithm specifically designed to reconstruct user-intended manifold surfaces from dense ribbon-format 3D brush strokes. This contribution is made possible by our detailed analysis of brush drawing characteristics.

\section{Previous work}
\label{sec:prev}
\label{sec:related}

Our work builds upon prior research \old{on sketch-based modeling, 3D curve drawing,  VR modeling interfaces, and surface reconstruction.}
\revised{across multiple domains.}

\paragraph{2D-Sketch-Based Modeling}
2D-sketch-based modeling methods infer depth information from collections of sparse 2D artist strokes, which are assumed to employ a specific drawing style and capture key properties \revised{of} the artist-intended shape~\cite{OSSJ09}. SurfaceBrush recovers surface geometry from dense 3D strokes, an input that exhibits very different properties.
Algorithms that process 2D drawings frequently leverage established drawing conventions and observations about human perception of 2D imagery~\cite{Xu:2014,Bae:2008,Nealen:2007,Schmidt:2009,SBSS12,Li:2017}. Such resources are essentially non-existent for 3D drawings, since \revised{until} \old{till} recently there had been few opportunities for artists to 
use 3D strokes to depict shape. We derive the characteristics of the inputs we seek to process via examination of publicly available VR artwork databases~\cite{TiltBrushRepo,TiltBrushRepo1} and an observational study of VR 3D shape drawing (Sections~\ref{sec:data},~\ref{sec:user_study}).

\revised{
\paragraph{Sketch Consolidation} Our work has conceptual similarities to sketch consolidation~\cite{Noris:2012,liu2018strokeaggregator,Liu:2015,Stahovich:2014,Fu:2011,Xing:2014}. However, in 2D each stroke vertex has unique nearest left/right neighbors along the stroke’s orthogonal. This property no longer holds in 3D, making determination of best pairwise vertex matches a lot more challenging.
}

\paragraph{3D Curve Drawing}
Researchers \revised{have} \old{had} proposed  a range of 
tools for creating, rendering, and manipulating  curves directly in 3D space~\cite{israel09,Grossman:2002,tano03,Jackson:2016,diehl04,amores17,Kim:2018}. 

Recent systems render captured curves in real time using head-mounted displays, depicting them  as ruled surface ribbons~\cite{Keefe:2001,Keefe:2007,TiltBrush} or as tubular shapes with cylindrical profiles~\cite{PaintLab,Keefe:2007}.

Utilizing the content artists produce using such systems for shape modeling requires converting raw curve drawings into 3D surface models. 
SurfaceBrush achieves this goal using as input oriented ribbon strokes created with the widely available TiltBrush system; it can also be employed in conjunction with other VR systems which support such strokes.

\paragraph{VR Modeling Interfaces.}
Researchers have explored a range of VR modeling interfaces.  VR sculpting tools 
\revised{\cite{Kodon,ShapeLab,OculusMedium:2016}} 
allow expert users to create sophisticated shapes. 
 VR interfaces that support Boolean operations over a fixed set of primitives~\cite{Tano:2013,GoogleBlocks,diehl04,DesignSpace} provide a promising avenue for modeling CAD geometries but are not well suited for free-form shapes. 
Others enable users to draw a range of swept surfaces in 3D space~\cite{GravitySketch,Keefe:2001,schkolne01,SchkolneTR}. 
To model complex shapes using this approach, users need to mentally break them into coarse, non-overlapping, sweepable patches, and separately draw each patch: a task that requires modeling expertise and is especially challenging for organic shapes. 
Several VR systems facilitate editing of existing  3D surfaces~\cite{Wesche:2001,Kwon:2005,GravitySketch}.
Our work complements all those systems in its focus on providing experts and amateurs alike with the means to author free-form manifold geometries which they can later edit. 

Several VR interfaces allow users to connect 3D curves into cycles or curve networks~\cite{Jackson:2016,Wesche:2001,Kwon:2005,fiorentino02} and provide  them with the option to surface those inputs using traditional cycle and network surfacing techniques, such as Coons patches or NURBs. The obtained surfaces are highly dependent on the choice of the surfacing method.   
Grossman~\shortcite{Grossman:2002} and Sachs~\shortcite{Sachs91} facilitate tracing of characteristic surface curves such as flow lines in a VR environment. Networks consisting of such curves can  be surfaced  using designated algorithms~\cite{Bessmeltsev:2012,Pan:2015}.
Employing any of these systems users need to understand the underlying surfacing method in order to draw the curves that would form their desired output.

Our approach does not require such understanding and does not constrain users to modeling particular surface families. As such we add another tool to the VR modeling palette, one specifically suited for non-expert users and generic free-form geometries.   

Our choice of using dense ribbon strokes as modeling input is inspired by the method of  Schkolne et al.~\shortcite{schkolne01,SchkolneTR} which forms free-form surfaces by merging adjacent swept surfaces drawn by artists. Schkolne et al. 
generate the merged surface  
using a method that is designed to provide a real-time approximation of point-cloud reconstruction techniques such as Alpha-Shapes~\cite{Edels}, Figure~\ref{fig:compare}. As the authors acknowledge, even on the relatively clean data they tested, the method frequently produces non-manifold geometries.

\paragraph{Surface Reconstruction from Curves, Point Clouds\revised{, and Triangle Soups}.}
Research on surface reconstruction from curves targets specific input sources and leverages their distinct properties. 
Many methods address reconstruction from closed, planar cross-section curves, e.g  \cite{Sharma:2016,Zou,Huang:2017}. Others address lofting, or surfacing of closed curve cycles~\cite{Gao2005,Varady2011,Schaefer2004,Nasri2009,Finch2011} and networks 
\revised{\cite{Bessmeltsev:2012,Pan:2015,Grimm:2012,Abbasinejad:2012,Wang:2016}}.
Our inputs do not conform to the assumptions employed by any of these methods: the strokes are not closed, are frequently non-planar, and do not form cycles or networks. They thus require a different set of priors for successful surfacing. 
\revised{Usumezbas et al. \shortcite{Usumezbas:2017} use curves on the surface of the output models while utilizing image data to filter out poor surfacing choices based on occlusions; we must process curves that extend inside the intended shapes, making occlusion a problematic criterion.}

Methods for surface reconstruction from point clouds~\cite{Berger:2017} can potentially be applied as-is to stroke polyline vertices or to a dense set of points sampled along the stroke ribbons (Figure~\ref{fig:compare} (b,c)). 
\setlength\columnsep{5pt}
\begin{wrapfigure}[]{l}[0pt]{.45\linewidth}
	\vspace{-10pt}
  \begin{center}
    \includegraphics[width=\linewidth]{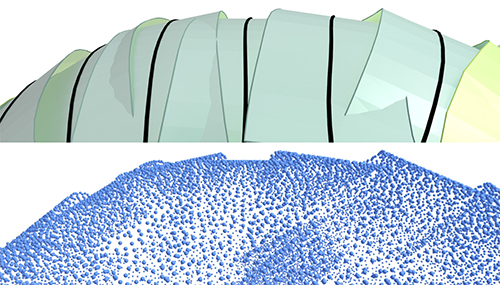}
  \end{center}
  \vspace{-5pt}
  \caption{Close-up of ribbons and ribbon samples on the teapot}
  \label{fig:stroke_zoom}
  \vspace{-10pt}
\end{wrapfigure}
However sampling brush stroke \revised{drawings} \old{drawing}  (Figure~\ref{fig:challenges}) produces point clouds with inconsistent normal orientation, multiple samples in the interior of the intended shape, and other artifacts inconsistent with the assumptions made by typical reconstruction techniques~\cite{Berger:2017}. Moreover, while stroke vertex locations are typically reflective of the intended surface location, due to the inaccuracy in the stroke normals, points sampled along the ribbons are often misplaced with respect to this surface (Figure~\ref{fig:stroke_zoom}).
These artifacts cause  traditional reconstruction methods, \revised{such as ~\cite{Kazhdan:2013,Edels,Bernardini:1999,Avron:2010,Xiong:2014,Wang:2016}} to fail dramatically (Figure~\ref{fig:compare}, Section~\ref{sec:results}).

\revised{One could potentially treat the brush strokes as triangle strips, and use methods for triangle soup surfacing and repair to attempt to recover the 3D shapes from them. However, voxel based methods, e.g. ~\cite{Ju:2004,Shen:2004}, are only applicable to inputs one expects to be closed. Roughly one quarter of our inputs have some open surface
elements (e.g. ground on the bonsai or feet on the chicken).  Even on closed surfaces, these methods fail drastically in terms of the topology and geometry of  the results produced (Section~\ref{sec:results}). Winding-number based approaches~\cite{Barill:2018} produce similar artifacts (Section~\ref{sec:results}).}

\begin{figure}
  \centering
  \includegraphics[width=\columnwidth]{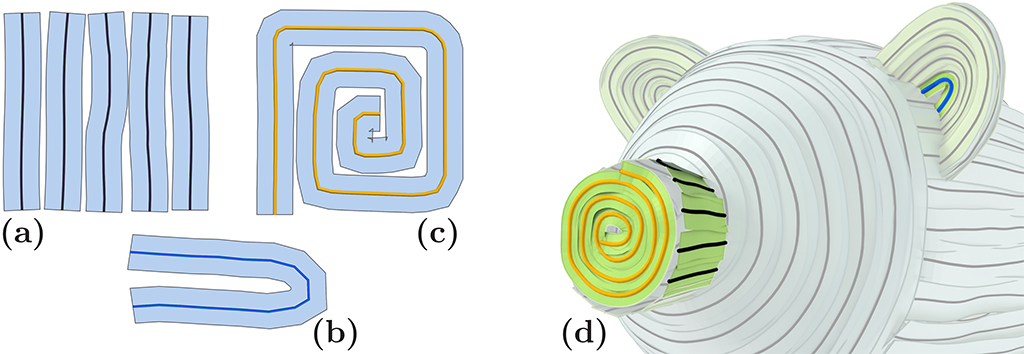} 
  \caption{Artists use different stroke patterns with locally similar tangents to cover, or paint, different surface regions: (a-c) frequent schematic patterns, (d) portion of a typical drawing where all three patterns are used. Piggy bank: $\copyright$ Elinor Palomares.}
  \label{fig:Patterns}
\end{figure}

\section{Input Drawing Characteristics}
\label{sec:input}
\label{sec:data}

Analysis of  publicly available VR artwork~\cite{TiltBrushRepo,TiltBrushRepo1} and observation of VR 3D shape drawings created by our study participants (Section~\ref{sec:user_study}) point to a number of core common characteristics  of 3D brush-stroke drawings. 
\\
{\em Dense coverage:} 
In both datasets, the drawn stroke ribbons frequently overlap and  typically densely cover the communicated shapes leaving relatively small inter-stroke gaps or holes whose size is typically smaller than the width of the surrounding strokes (e.g. Figure~\ref{fig:teaser}). The stroke width users employ varies across different parts of the surface, and is typically more narrow on finer features. \\
{\em Local tangent consistency:} Artists frequently draw contiguous surface patches using strokes with similar tangent directions (Figures~\ref{fig:challenges},~\ref{fig:Patterns}).  Tangent consistency is local rather than global since artist often use very different stroke directions in different parts of the model: they choose stroke directions based on drawing convenience and often align stokes with the minimal absolute curvature directions on the intended surface.  \\
{\em Persistent adjacency:} Artists use a range of drawing strategies when forming tangent-consistent stroke patches: they may use multiple side-by-side strokes, draw sharply-turning self-adjacent strokes, or use long self-adjacent spirals (Figure~\ref{fig:Patterns}). The strategy may often vary across a single input. Adjacent side-by-side tangent consistent strokes typically have comparable lengths. Consequently most  strokes have only a few, and often just one, immediately adjacent, similarly-directed strokes on each side.\\
{\em Normal orientation:} The input stroke normals are defined by the orientation of the hand-held controller. 
Users typically aim for the stroke ribbons to lie in the tangent plane of the intended surface, thus the stroke normals are typically roughly orthogonal to this surface (Figures~\ref{fig:teaser},~\ref{fig:challenges}) but are rarely exact.  
\revised{VR systems use double-sided ribbon rendering, which obscures stroke orientations from artists. Consequently, we observe that artists do not attempt any in-out consistency, producing strokes whose orientation is essentially a function of drawing access convenience. Specifically,} users typically hold the controller like a brush, with its  tip pointing away from them, resulting in stroke normals that point toward the artist much of the time. Consequently, normal direction is determined by the location of the artist relative to the drawn shape and is typically {\em not} reflective of the surface  front-back orientation (Figure~\ref{fig:challenges}).
\revised{As this figure illustrates, normal mis-orientation is a persistent feature. Approximately one-third of the strokes in our inputs are oriented in the opposite direction to the plurality. This ratio holds across artists and input categories. Thus orientation inconsistency must be addressed by any method processing VR-brush strokes.}  \\
{\em Intersecting and isolated strokes:} When drawing different model parts, artists rely on the what-you-see-is-what-you-get principle and assume that making strokes or portions of strokes not visible from outside the object is tantamount to erasing them. Thus, when drawing different parts of the target shape, they often extend strokes into the interior of the models producing multiple intersecting stroke groups (Figures~\ref{fig:challenges},~\ref{fig:components}) and do not erase occluded outlier strokes. Both existing artwork and 3D drawings created at our bequest often use sparse,  isolated, strokes for communicating one-dimensional or very narrow geometries (such as \revised{the} chicken feet in Figure~\ref{fig:challenges}). We speculate that this choice \old{reflect} \revised{reflects} the difficulty of accurately depicting outer surfaces of narrow features and leverages the fact that human observers can easily parse such abstracted or skeletal elements.\\
{\em Stroke Accuracy:} Lastly, we note that users aim to accurately communicate the envisioned shape, thus the shape and location of most stroke polylines typically reflects the intended surface geometry along them up to low-frequency noise inevitable when drawing 3D  content by hand. Note that the accuracy of any \old{points} \revised{point} on the ribbons away from the polyline depends on both the accuracy of the stroke normals and the underlying surface curvature - when the curvature in the direction orthogonal to the strokes is large, even with perfect normals, ribbon sides can significantly deviate from the surface (Figure~\ref{fig:stroke_zoom}).

\begin{figure}[h]
\vspace{-3pt}
\includegraphics[width=\columnwidth]{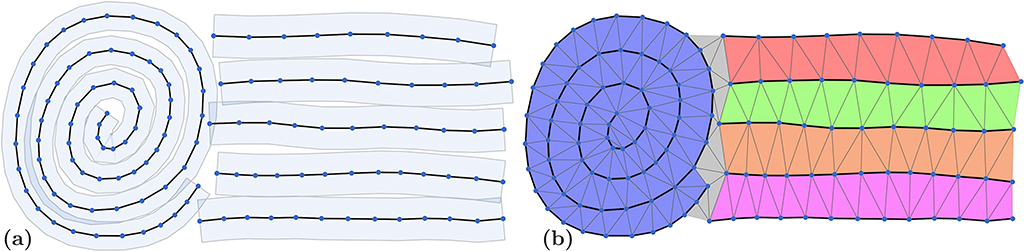} 
\caption{Schematic surfacing illustration: (a) input strokes, (b) output mesh consisting of inter-stroke mesh strips bounded by adjacent stroke polylines (multicolor) computed first, and gap spanning strips (gray) computed later.}
\vspace{-7pt}
\label{fig:schematic}
\end{figure}

\begin{figure*}[h]
  \centering
  \includegraphics[width=\textwidth]{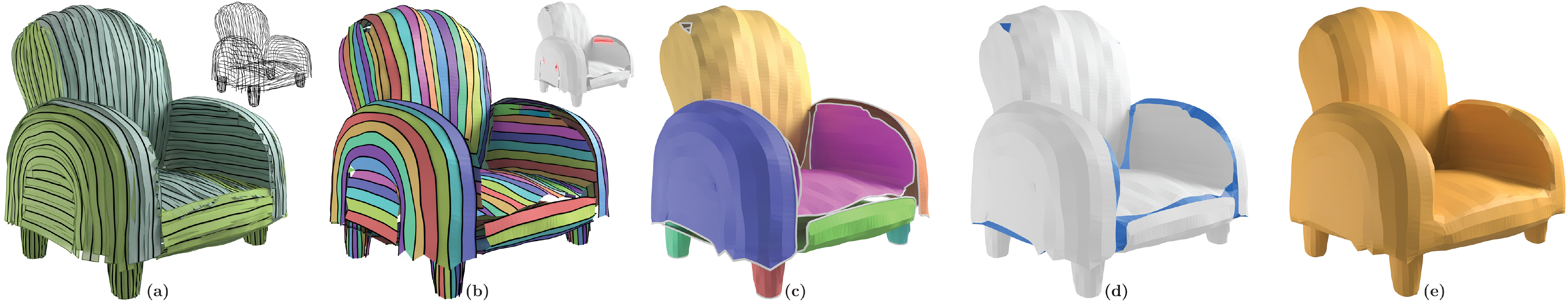} 
  \caption{SurfaceBrush stages: (a) Input stroke polylines (black) and ribbons (color reflects normal orientation). (b,c) Inter-stroke strip computation: initial mesh strips imposed by vertex-to-vertex matches (non-manifold edges highlighted in the inset) (b) and output manifold partial surfaces (each connected component rendered in a different color) (c). Gap closure (d,e): gap-spanning mesh strips (d), and final surface (e). Input drawing: $\copyright$ Elinor Palomares.}
  \label{fig:overview}
\end{figure*}

\section{Overview}
\label{sec:overview}

The SurfaceBrush framework is designed to operate in conjunction with existing 3D stroke drawing tools and to process as input completed ribbon stroke drawings depicting manifold, open or closed, surfaces.  It converts these drawings into manifold surface meshes describing the artist-intended geometry (Figure~\ref{fig:overview}).

\subsection{Surfacing Goals}
The observations about the key properties of 3D VR drawings (Section~\ref{sec:data}) allow us to formulate our algorithm's goals in terms of producing outputs consistent with user expectations. 
\\
{\em Interpolation and Normal Consistency:}
While we expect some strokes or stroke sections to be outliers, we expect most stroke edges to be part of the target surface and to accurately depict its location.  
Thus we expect the reconstructed surface to {\em interpolate} the vast majority of stroke edges and expect this surface to be roughly orthogonal to the interpolated stroke vertex normals. \\
{ \em Union of Inter-Stroke Strips:} Combining these expectations with observations about tangent consistency and persistence we argue that users expect the interpolating surface to be dominated by {\em surface strips} connecting, or {\em bounded} by, side-by-side stroke sections with similar tangent directions (Figure~\ref{fig:schematic}b, multicolor). Each such strip consists of a sequence of triangles where each triangle shares one edge with its predecessor. Jointly, these strips form one or more {\em partial surfaces} interpolating most stroke vertices and edges (Figure~\ref{fig:schematic}b has two such surfaces separated by the gray strip). Since we expect the adjacencies between strokes to be persistent, we expect the number of different strips bounded by each given stroke to be small (frequently just one on the left and one on the right).  \\
{\em Gap Closure:} We expect the final surface to connect the partial surfaces closing the gaps between them. As before, we expect these {\em gap-spanning} surface strips (Figure~\ref{fig:schematic}b gray) to connect close-by boundary vertices and to be orthogonal to the partial surface normals at these vertices. \\
\setlength\columnsep{1pt}
\begin{wrapfigure}[7]{l}[0pt]{0.25\linewidth}
  \vspace{-10pt}
  \begin{center}
    \includegraphics[width=\linewidth]{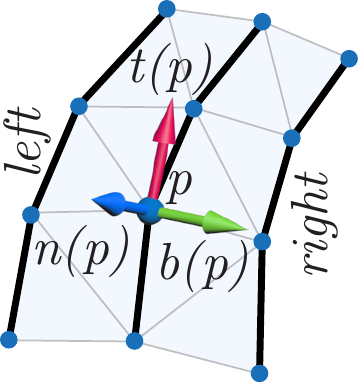}		
  \end{center}
\end{wrapfigure}
{\em Manifoldness:} To satisfy manifoldness, each stroke section must bound at most two surface strips. Since we expect the stroke normals to be orthogonal to both strips and expect the output surface to be fair, one of these strips should be on the right and the other on the left of the section with respect to the local Frenet frame defined by the stroke tangent and normal (see inset).  We expect sections along the 
partial surface boundaries to bound at most one gap-spanning strip located on the opposite side of the partial surface with respect to a Frenet frame defined by the boundary tangents and the partial surface normals along them. 

\begin{figure}
 \includegraphics[width=\columnwidth]{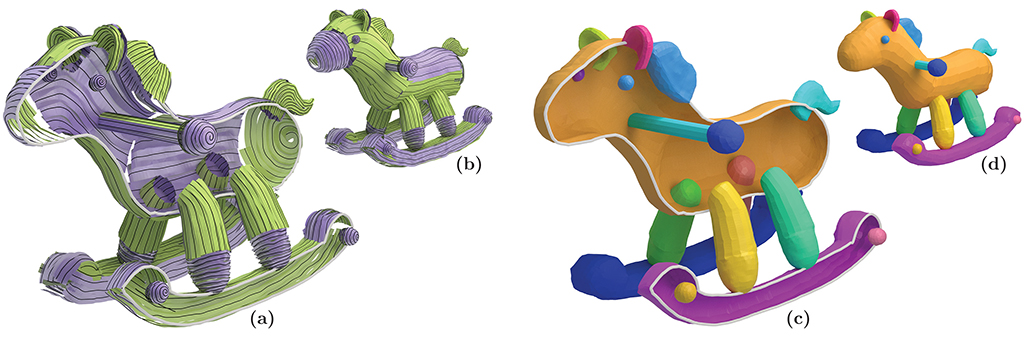} 
\caption{(a) The horse (Figure~\ref{fig:teaser} contains multiple disjoint intersecting stroke groups (see cutout view), (b) resulting surfacing output with each connected component drawn in separate color. We use a Boolean union of these parts as input to fabrication (Figure~\ref{fig:teaser}d). Input drawing: $\copyright$ Jafet Rodriguez.}
\label{fig:components}
\end{figure}

We can thus formulate our overall surfacing goal as generating a manifold union of inter-stroke and gap-spanning strips that interpolate the vast majority of the input stroke edges and vertices (Figure~\ref{fig:schematic}, right). The inter-stroke strips need to connect side-by-side stroke sections, and all strips need to be persistent and connect adjacent stroke vertices with similar normals (up to orientation). By design, we do not seek to connect distinctly separate connected components (Figure~\ref{fig:components}), leaving this optional step to the user.

\revised{Notably, relaxing the manifoldness constraint makes the problem much easier. However, non-manifold meshes cannot be processed by many mesh processing algorithms and are not supported by many commonly used data-structures, making the results significantly less usable.}

\subsection{Algorithm}
\label{sec;algo}

We \old{design} \revised{designed} our algorithm based on the requirements above. 
Since the geometry of the gap-spanning mesh strips can only be determined once all inter-stroke strips are in place, we compute the inter-stroke mesh strips first (Section~\ref{sec:local}) and then compute the gap-spanning ones (Section~\ref{sec:global}). \revised{This separation into stroke and gap surfacing steps allows us to take advantage of the directional similarity between strokes first, and to subsequently leverage direction similarity between (previously non-existent) partial-surface boundaries.}
Our first step computes dense matches between stroke vertices, then uses 
these matches to form initial mesh strips between the strokes (Figure~\ref{fig:overview}c) and finally removes the non-manifold artifacts in the resulting mesh   (Section~\ref{sec:local}). 
SurfaceBrush employs a similar \old{three step} \revised{three-step} solution process during the gap processing step to match and then connect the boundaries of the partial surfaces using gap-spanning mesh strips (Section~\ref{sec:global}).

\paragraph{Pre-Processing}
When artists use digital sketching tools, they often activate the stylus or controller trigger a few milliseconds before starting the stroke drawing motion and \old{disactivate} \revised{deactivate} it a few milliseconds after concluding the motion~\cite{liu2018strokeaggregator}. This behavior produces short randomly oriented stroke sections next to stroke end-points. Our pre-process removes these redundant sections using an approach similar to \cite{liu2018strokeaggregator}: we check if the strokes have an abrupt direction change (angle of $45^\circ$ or less between consecutive tangents) within 15\% of overall stroke length from either end  and remove the offending end-sections.


\section{Inter-Stroke Surface Strips}
\label{sec:local}

At the core of our framework is the need to match sections, or edge sequences, along input strokes that bound surface strips on the artist-envisioned surface.

\setlength\columnsep{1pt}
\begin{wrapfigure}[]{l}[0pt]{8mm}
  \vspace{-10pt}
  \begin{center}
    \includegraphics[width=6mm]{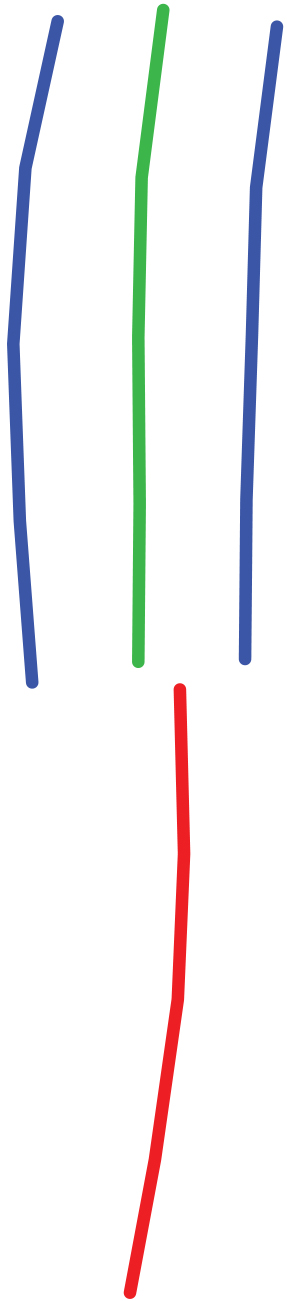}
  \end{center}
  \vspace{-10pt}
\end{wrapfigure}
 When matching stroke sections, we seek matches that reflect four key properties: proximity, tangent similarity, persistence, and normal consistency. 
Since we seek a manifold output, we expect each stroke section to have at most one matching section on its left and one on its right.

\begin{figure}
\centering
\includegraphics[width=\columnwidth]{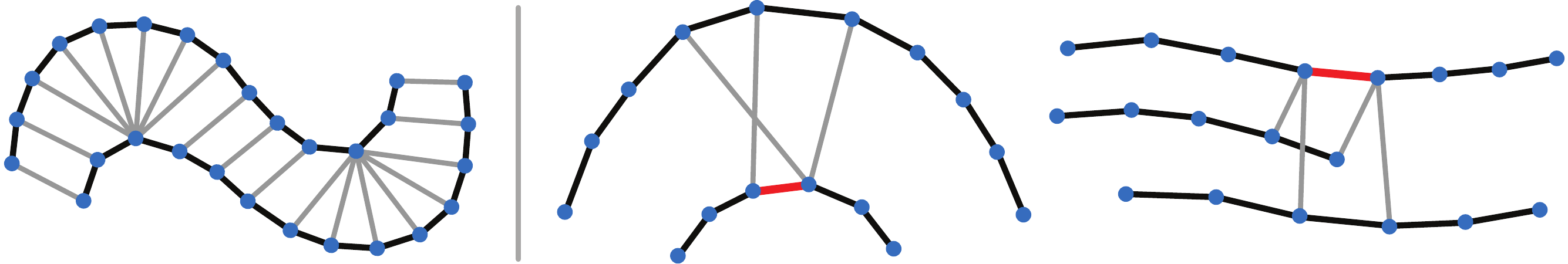} 
\caption{(left) Desirable non-bijective matches accounting for uneven density. (right) Groups of matches that induce non-manifold meshes. Note that one cannot determine if the result will be manifold by looking at a subset of the matches (pairwise matches in 3D that cross one another in some view do not necessarily induce incompatible matches.)}
\vspace{-10pt}
\label{fig:manifold}
\end{figure}

Sections in the middle of a cluster of side-by-side strokes should have matches on both sides (inset, green); sections along the boundaries of such clusters should have a matching section only on one side (inset, blue); and outlier or isolated sections should have no matches on either side (inset, red). 
The partition of strokes into sections and the classification of  these sections into the three types above are not known a priori and need to be deduced by our algorithm.  We simultaneously segment strokes into sections and match them with respective sections on the same or other strokes, using a discrete optimization framework, that operates on the stroke's vertices. Specifically, we first obtain pairwise vertex-to-vertex matches and then use those to obtain the stroke sections and the correspondences between these sections: each pair of matching sections is defined by a maximal consecutive sequence of vertices on one stoke that match to another consecutive vertex sequence on the same or other stroke. Note that the vertex-to-vertex matches should not necessarily be bijective - given strokes with different vertex density
we want to allow many to one matches to enable dense correspondences (Figure~\ref{fig:manifold},~left).

To account for the demands above, we need to obtain vertex-to-vertex matches that satisfy three types of criteria: (1) criteria that can be assessed at the level of individual pairwise vertex matches, (2) criteria that require assessing  two matched vertex pairs at once, and (3) criteria that require assessing three or more pairs in tandem. Specifically, proximity, tangent similarity, and normal consistency can be assessed at the individual vertex-to-vertex match level.
Promoting persistence implies prioritizing configurations where consecutive stroke vertices match to similarly consecutive vertices, necessitating assessing two matched vertex pairs at once.
Lastly, assessing manifoldness requires analyzing, and consequently disallowing, configurations of three or more matched pairs (Figure~\ref{fig:manifold}, right),  as smaller subsets do not necessarily provide sufficient information.

Even a simpler variant of our problem, one where the decision about incompatible matches can be done by assessing two matched pairs (rather than a larger group) is shown to be NP-complete via a reduction from 3D matching, which was shown to be NP-complete by Karp~\shortcite{Karp72}. The reduction is straightforward: the pairs in this problem correspond to sets in the 3D matching problem, and two pairs are prevented from coexisting unless the corresponding sets are disjoint. 
Thus, obtaining matches that satisfy our criteria using \old{off-the shelf} \revised{off-the-shelf} methods is impractical.

\begin{figure}
 \includegraphics[width=\columnwidth]{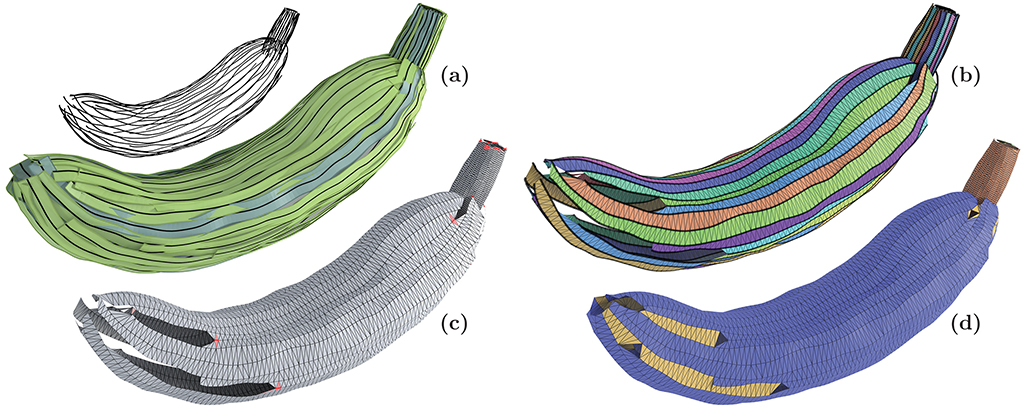} 
\caption{Inter-stroke mesh strip formation: (a) input strokes and ribbons (inset shows stroke polylines) (b) initial mesh strips reflecting restricted matches (c) non-manifold edges and vertices in the initial mesh highlighted in red, (d) output partial meshes after consolidation and extension. Input drawing: $\copyright$ Jafet Rodriguez.}
\vspace{-10pt}
\label{fig:inter_stroke}
\end{figure}

We nevertheless efficiently obtain a desirable solution that accounts for all three criteria types by using a multi-stage matching method that leverages the anticipated persistence of the matches we seek (Figure~\ref{fig:inter_stroke}).  

\setlength{\belowcaptionskip}{-12pt}
\setlength{\belowcaptionskip}{0pt}

We first note that absent the manifoldness requirement, the matches we seek for can be computed independently for each stroke.  
Specifically, for a single stroke we can cast the optimization of the remaining criteria as a maximization of  a score function (Section~\ref{sec:asses}) that accounts for both the quality of individual matches and for persistence, or pairwise compatibility between the matches at consecutive stroke vertices. The matches that maximize this combined function can be efficiently computed using a classical dynamic programming framework~\cite{Viterbi} (Section~\ref{sec:asses}). This method, however, is designed for finding matches for all vertices and consequently cannot account for cluster borders or outliers. Thus, to avoid undesirable matches, we restrict the set of per-vertex {\em matching candidates} during this computation. Our first matching pass (Section~\ref{sec:match}) uses very conservative matching candidate sets, generating correct matches for a large subset of vertices but intentionally leaving some vertices unmatched. 
We use the computed matches to define inter-stroke mesh strips (Section~\ref{sec:mesh}, Figure~\ref{fig:inter_stroke}b). 

We eliminate non-manifold configurations in the resulting mesh (Figure~\ref{fig:inter_stroke}c) while minimally reducing the matching score function by formulating these goals as a classical correlation clustering  problem~\cite{bansal2004correlation}  and solve it using an approximation method  (Section~\ref{sec:global_filtering}). While the problem solved in this step remains NP-hard, 
thanks to our restrictions on the possible matches assessed and our enforcement of persistence between the matches, the number of non-manifold artifacts in the resulting mesh is very small. Thus, they can be efficiently and effectively resolved by applying the clustering to only small subsets of the mesh triangles enabling speedy solution.

The restrictions on the matching candidate sets imposed in our first matching pass (Section~\ref{sec:match}) may result in unmatched stroke vertices for which suitable matches do exist (the unfilled spaces between strokes in Figure~\ref{fig:inter_stroke}b). We generate mesh strips connecting such previously unmatched vertices by applying the matching, meshing and manifold consolidation steps again to stroke sections along the boundaries of the current partial surface, using an updated more lax matching candidate set (Section~\ref{sec:local_repeat}). The output of this step is a manifold partial surface mesh interpolating stroke groups with similar directions (Figure~\ref{fig:inter_stroke}d).

\subsection{Match Computation}
\label{sec:asses}

When looking for matches, we distinguish between left and right sides of each stroke using the direction of the stroke binormal 
$b(p) = t(p) \times n(p)$ in the local Frenet frame at each stroke vertex $p$ defined by the stroke's tangent $t(p)$ and normal $n(p)$, to distinguish between these sides.  Given a stroke $S = p_0,\ldots ,p_n$ and a candidate set of matching vertices for each stroke vertex $p$, we evaluate the potential left (or right) matches $(p_i,q_i)$ using a combination of 
vertex-to-vertex scores $S^v_l(p_i,q_i)$ for left side matches and $S^v_r(p_i,q_i)$ for right-side matches,
and a persistence score $S^e(p_i,p_{i+1},q_i,q_{i+1})$ that assesses the compatibility between the potential matches of consecutive stroke vertices.
  
Both scores are designed to be symmetric so as to prioritize matches which are bijective whenever possible, in order to reduce the occurrence of non-manifold artifacts. We define the combined score of matching the vertices of $S$ to the vertices $Q=q_0,\ldots, q_n$ as their left side matches as 
\begin{equation}
M_l(S,Q) = \prod_i S^v_l(p_i,q_i) S^e(p_i,p_{i+1},q_i,q_{i+1})
\label{eq:stroke}
\end{equation}
We define $M_r(S,Q)$ in a symmetric manner and look for left and right matches that maximize these scores. 
We use a product rather than a sum to discourage outlier matches. Maximizing the per-stroke scores can be seen as a variant of the classical Markov chain optimization.  Given a set of {\em matching candidates} $C(p_i)$ for each vertex $p_i \in S$, we can compute the matches within these sets that independently maximize $M_l(S)$ or $M_r(S)$ in polynomial time using the classical Viterbi algorithm~\cite{Viterbi}.
To obtain a valid solution, we exclude from the per-stroke scores vertices with empty matching candidate sets $C(p_i)$ or the edges emanating from them. 
The strategy we employ to compute the matching candidate sets during different stages in our surfacing process is elaborated on in relevant sections below. 
We define our overall matching goal  as maximizing the matching scores across all strokes $S$ in our drawing,
\begin{equation}
M = \sum_S M_r(S,Q)+M_l(S,Q).
\label{eq:global_score}
\end{equation}

 Absent any constraints, this goal can be achieved by  maximizing the scores for each  stroke individually (as there is no requirement for the matches to be symmetric).

\begin{figure}
\includegraphics[width=\columnwidth]{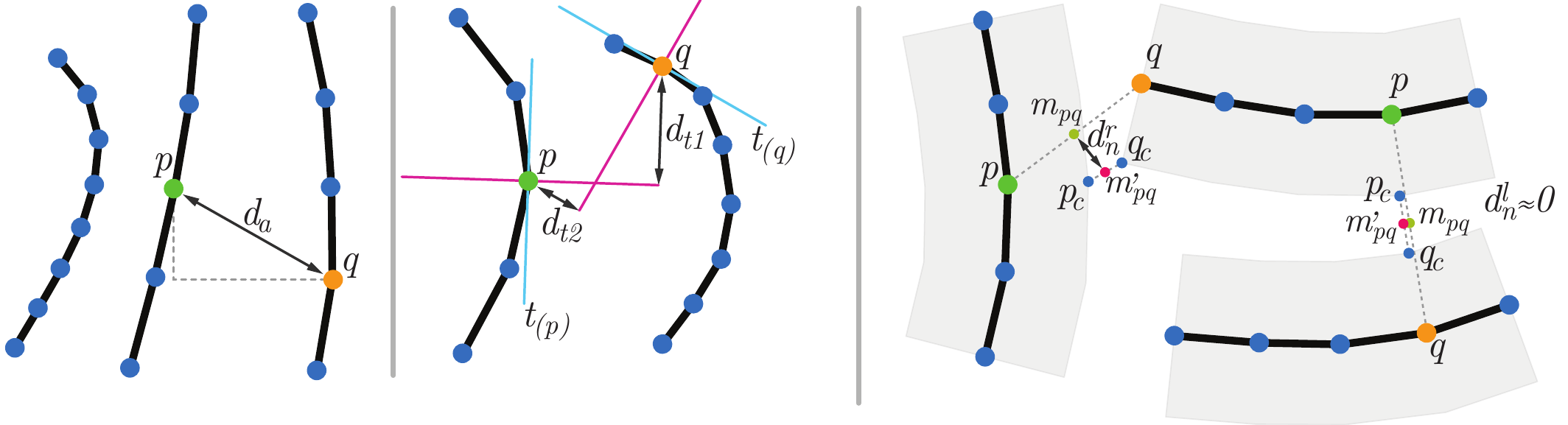} 
\caption{Components of the vertex matching score $S^v_r(p,q)$: $d_a$, $d_t$, $d^r_n$}
\label{fig:individual}
\end{figure}
\paragraph{Vertex-to-Vertex Matching Score}

Given a pair of vertices $p$  and $q$, we define the score of using $q$ as the left or right match of $p$ as a function of three distance terms, designed to be on the same scale (Figure~\ref{fig:individual}). 
 The first is the absolute distance between them
\[ d_a = \|p-q\|.\]
The second term measures the degree to which the vertices are side-by-side with respect to their respective tangents, $t(p)$ at $p$ and $t(q)$ at $q$, as 
\[d_t =\frac{|(p - q) \cdot t(p)| + |(p - q) \cdot t(q)|}{2}.\]
We set $t(o) = (o_n - o_p) / \|o_n-o_p\|$ where $o_n$ and $o_p$ are the next and previous vertices on the stroke of the vertex $o$.

Lastly, we use the following construction to measure the degree to which the vector $\overrightarrow{pq}$ is orthogonal to the stroke normals at $p$ and $q$ and to \old{asses} \revised{assess} whether the matches are consistent with respect to the strokes' Frenet frames, namely whether the left (or right) match of each vertex is on \old{it's} \revised{its} right side with respect to its Frenet frame

(see Figure~\ref{fig:individual}, right).
When assessing a left match, we compute 

an offset vertex $p_c$ located at distance $w(p)$ on the left side of $p$ along the frame's binormal $b(p)$, where $w(p)$ is the user-specified stroke width at $p$.   

We compute both left and right offset vertices $q_r,q_l$ for $q$ using a similar strategy and using offset magnitude $w(q)$. 
 We set $q_c$ to the offset vertex of $q$ closest to $p_c$ and compute their midpoint $m'_{pq} = (p_c+q_c)/2$. When $q$ lies to the left of $p$ and $\overrightarrow{pq}$ is orthogonal to the stroke normals at $p$ and $q$, this midpoint \revised{$m'_{pq}$} coincides with the midpoint $m_{pq} = (p + q)/2$.  If either one of the criteria does not hold, the two midpoints will be far apart (see Figure~\ref{fig:individual},right). Following this observation, we define
\[d^l_n = \| m_{pq} - m'_{pq}\|.\]

We define the \old{overal} \revised{overall} score for assigning $q$ as the left match of $p$:
\begin{equation}
  S^v_l(p,q) = \rm{e}^{\frac{-(d_a + d_t + d^l_n)^2}{2 \sigma^2}}.
  \label{eq:Emission}
\end{equation}
We define the right-side assignment score $S^v_r(p,q)$ using $d^r_n$ computed symmetrically to $d^l_n$. 

Our empirical observations indicate that users rarely leave unintentional gaps between side-by-side stroke ribbons that are wider \old{that} \revised{than} half of these strokes' widths. 
Thus, we expect the values of each of the three distance metrics $d_a$, $d_t$ and $d_n$ for most desirable matches to be less than $d_{max}=1.5 (w(p)+w(q))/2$.  
Consequently, we expect pairwise matches where the sum of the three distances exceeds $3 d_{max}$ to be undesirable. 
Using the three sigma rule we encode this preference by setting $\sigma = d_{max}$ .

\paragraph{Persistence Score}

\setlength\columnsep{1pt}
\begin{wrapfigure}[]{l}[0pt]{0.17\linewidth}
  \vspace{-10pt}
  \begin{center}
    \includegraphics[width=\linewidth]{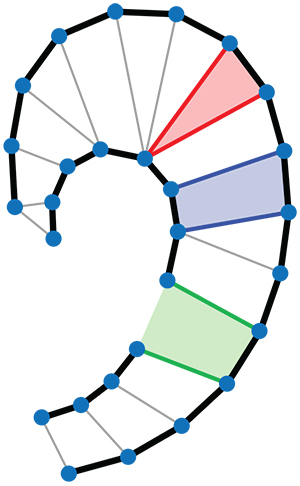}
  \end{center}
  \vspace{-10pt}
\end{wrapfigure}
Persistence requires the majority of consecutive vertices along a given stroke to match to similarly consecutive vertices (blue in inset). Exceptions include \old{discretizaton} \revised{discretization} mismatches (red in inset) and transitions between stroke sections that bound different strips (green in inset). We account for persistence without unduly penalizing such exceptions and assess the acceptability of these exceptional cases when they occur by formulating this score using geometric rather than topological properties. 
In addition to promoting persistence, the score we use further reinforces our preference for matching side-by-side, parallel, stroke sections. 

Given a pair of consecutive vertices $p_i,p_{i+1}$ that match to a pair of vertices $q_i$ and $q_{i+1}$, respectively, we measure persistence using a combination of three distances
\begin{eqnarray}
d_p = \|(p_{i+1} - p_i) - (q_{i+1} - q_i)\|+ \|(p_{i+1} - q_i) - (q_{i+1} - p_i)\|+ \nonumber \\
 \|(p_{i+1} - q_{i+1}) - (p_i - q_i)\| \nonumber
\end{eqnarray}
The first term promotes matches \revised{that} \old{which} have the same spatial relationship between the edge $p_i,p_{i+1}$ and the line $q_i,q_{i+1}$. The second and third jointly promote co-planarity and parallelism between them. \revised{These terms zero out when the edges are both parallel and coplanar and jointly reflect how far they are from satisfying these conditions.}

We convert this distance sum into a score in the $[0,1]$ range as follows.  
\begin{equation}
  S^e(p_{i,i+1},q_{i}|q_{i+1}) = \rm{e}^{\frac{-d_p^2}{2\sigma^2}.}
  \label{eq:Transition}
\end{equation}
We use the same value of $\sigma$ as for the vertex-to-vertex matching score, following the same argument.


\subsection{Restricted Matching}
\label{sec:matching}
\label{sec:match}

Directly computing the best matches for each stroke while including all vertices on  all strokes in the candidate sets of each 
vertex is computationally expensive. Moreover, our per-stroke score optimization is defined so as to find  left (right) matches for each vertex with a non-empty left (right) matching candidate set. 
Yet, user drawings may depict open surfaces whose boundary vertices should have matches only on one side, and isolated strokes which should have no matches on either side. 
Thus, to avoid outlier matches, we need to discard potential outliers during candidate matching set computation.   
Our restricted matching pass obtains conservative, reliable matches by leveraging our expectation of match persistence.
Persistence indicates that most strokes are likely to have just a few, and frequently only one matching stroke on the left or right. 
Following this observation, instead of looking for per-vertex matches globally, we first locate for each stroke a single most likely, or most \revised{{\em dominant}}, \revised{{\em neighboring}} stroke on its right and a single one on its left.  
We then compute the best left and right per-vertex matches along each stroke using a restricted set of matching candidates, which only includes vertices on these dominant neighboring strokes, if they exist, and on the currently processed stroke itself.

\paragraph{Locating Dominant Neighboring Strokes.}
To locate one dominant left and one dominant right neighbor for each stroke $S$, we first compute matches for vertices along this stroke that maximize $M_l(S)$ and matches that maximize  $M_r(S)$, \revised{ by} considering possible matching candidates on \revised{{\bf all}} strokes. \old{and} \revised{We then} use the frequency of matches \old{between them} \revised{from the stroke $S$ to other strokes} to \old{identify} \revised{define} the dominant \revised{left and right} neighbors \revised{for this stroke}.  

During this \revised{first matching pass} \old{initial match computation}, we \old{restrict} \revised{define}
 the left (right) {\em matching candidate set} for each vertex $p$ 
to include vertices $q$ across \revised{{\bf all}} input strokes that satisfy the following \revised{{\em baseline} matching} conditions: \old{(1) $ \|q-p\| \leq d_{max}$; (2) the angle between $q-p$ and the binormal $b(p)$ is at most $60^\circ$ (for the right candidate set we assess the angle between $q-p$ and $-b(p)$); (3) $q \neq p$ and $q$ is not an immediate neighbor of $p$ along its stroke. These restrictions reduce the likelihood of outlier matches and speed up computation.}

\revised{
\begin{enumerate}
\item{$ \|q-p\| \leq d_{max}$ }
\item{ the angle between $q-p$ and the binormal $b(p)$ is at most $60^\circ$ (for the right candidate set we assess the angle between $q-p$ and $-b(p)$)}
\item{$q \neq p$ and $q$ is not an immediate neighbor of $p$ along its stroke.}
\end{enumerate} 
The baseline matching conditions are designed to reduce the likelihood of outlier matches and to speed up computation by reducing the solution space.}

\setlength\columnsep{2pt}
\begin{wrapfigure}[8]{l}[0pt]{0.1\linewidth}
  \vspace{-10pt}
  \begin{center}
    \includegraphics[width=\linewidth]{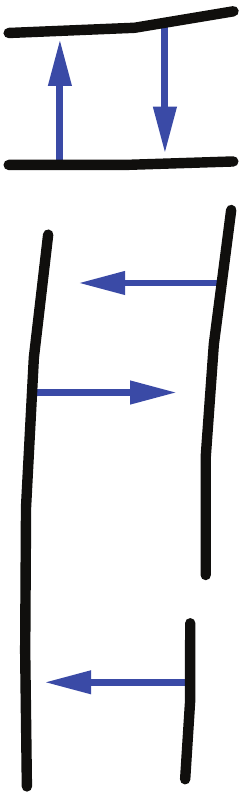}
  \end{center}
  \vspace{-5pt}
\end{wrapfigure}
We define the left (right) \revised{{\em matching frequency}} $F_{ST}$ from stroke $S$ to stroke $T$ as the percentage of vertices $p_i \in S $ that match vertices $q_j \in T$ as their left (right) matches. Note that this value is not symmetric: given for instance two side-by-side strokes where one is shorter than the other, the frequency for mapping the shorter to the longer will be higher than the other way around. 
We define a stroke \revised{$T$ to be the \textit{dominant left (right) neighbor of stroke $S$}} if the following three conditions hold: the left (right) matching frequency from $S$ to $T$ is higher than from $S$ to any other stroke; this frequency $F_{ST}$ is 
at least 30\%; and at least one pair of consecutive vertices on $S$ matches a pair of consecutive vertices on $T$ (the latter constraint discards T-junction matches where a stroke matches an end-vertex of another).  As the inset shows (arrows point to the computed dominant neighbors), we intentionally do not enforce symmetry in this process - allowing multiple strokes to share the same stroke as their \old{best left or right match} \revised{dominant left or right neighbor} and have strokes with no neighbors on one or both sides.

\paragraph{Restricted Matching Candidate Set.}
We define the left (right) restricted matching set of each vertex along a given stoke to  include vertices on the same stroke and its left (right) dominant neighbor, if one exists, that satisfy \revised{the baseline matching} conditions (1) to (3). We restrict this set further in the vicinity of stroke end-vertices, forcing 
condition (2) to hold at both $p$ and $q$. 
We use these restricted matching candidate sets  to compute the left and right per-vertex matches that optimize $M_l(S)$ and $M_r(S)$. Limiting the matching candidate sets drastically reduces the likelihood of outlier matches and produces locally optimal results along each individual stroke (Figure~\ref{fig:inter_stroke}b).

\subsection{Mesh Strip Generation}
\label{sec:meshing}
\label{sec:mesh}

Our meshing step receives as input a set of vertex-to-vertex matches $p_i,q_j$ between vertices on the same or different strips. 
It uses match pairs containing consecutive vertices on each stroke to 
determine the local meshing strategy (see inset). 
Given  two consecutive match pairs $p_i,q_j$ and $p_{i+1},q_{j+1}$ (or similarly $p_{i+1},q_{j-1}$) it triangulates the quad $p_i,p_{i+1}, q_{i+1}, q_i$  (or similarly $p_i,p_{i+1}, q_{i-1}, q_i$) using the diagonal that produces a more planar, better shaped triangulation.
\setlength\columnsep{1pt}
\begin{wrapfigure}[11]{l}[0pt]{0.23\linewidth}
  \vspace{-10pt}
  \begin{center}
    \includegraphics[width=\linewidth]{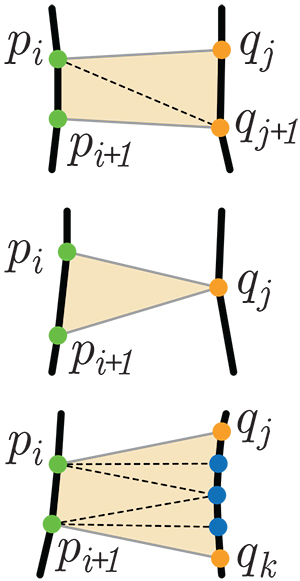}
  \end{center}
  \vspace{-10pt}
\end{wrapfigure}
Since we expect the mesh to be fair, it discards the quad if the dihedral angle between the resulting triangles is under $45^\circ$.

Given  two consecutive pairs $p_i,q_j$ and $p_{i+1},q_j$, it forms the triangle $p_i,q_j,p_{i+1}$.
Given consecutive vertices $p_i,p_{i+1}$ that match two non-consecutive vertices $q_j,q_k$ on the same stroke, it triangulates the polygon formed by the edges $(p_i,q_j), (q_{k},p_{i+1}),(p_{i+1},p_i)$  and the section $q_j,q_k$ \revised{only} if we have no matches between any pair of vertices within this section. This condition is used to avoid introducing non-manifold configurations. It triangulates the polygon using edges that connect interior vertices along the section  $q_j,q_k$ to $p_i$ or $p_j$, selecting a manifold, consistently oriented triangulation that maximizes the matching score along the section.

\begin{figure}
  \centering
  \includegraphics[width=\columnwidth]{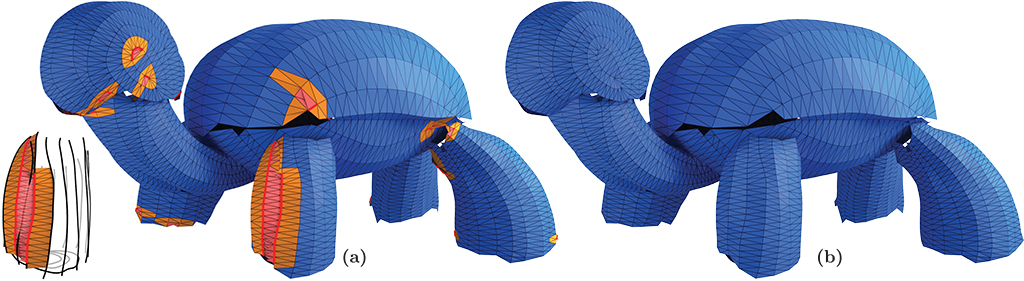} 
  \caption{(a) Annotated consolidation input:  incompatible triangles (red, offending edges highlighted ) undecided triangles (orange), unaffected ``output'' triangles (blue). (b)  Consolidated manifold mesh. Inset shows the strokes that trigger the non-manifold artifacts on the leg.}
  \vspace{-5pt}
  \label{fig:filtering}
\end{figure}

\subsection{Manifold Consolidation}
\label{sec:bijective}
\label{sec:filtering}
\label{sec:global_filter}
\label{sec:global_filtering}

We expect each stroke section \revised{in the final output mesh} to bound at most one mesh strip on its left and right.
Violating this expectation produces partial surfaces with non-manifold edges or vertices (Figure~\ref{fig:filtering}a). 
While our matching strategy is designed to minimize the likelihood of such  non-manifold artifacts, it does not fully prevent them. \revised{Thus the partial mesh defined by the union of mesh strips computed as described above may contain  non-manifold edges and vertices. Our manifold consolidation step removes a subset of the triangles surrounding such non-manifold entities to produce a manifold output mesh. In selecting the subset to remove it seeks to maintain as many triangles as possible in place, while optimizing the matching quality along mesh edges connecting matched vertices.}

\setlength\columnsep{2pt}
\begin{wrapfigure}[9]{l}[0pt]{0.15\linewidth}
  \vspace{-15pt}
  \begin{center}
    \includegraphics[width=\linewidth]{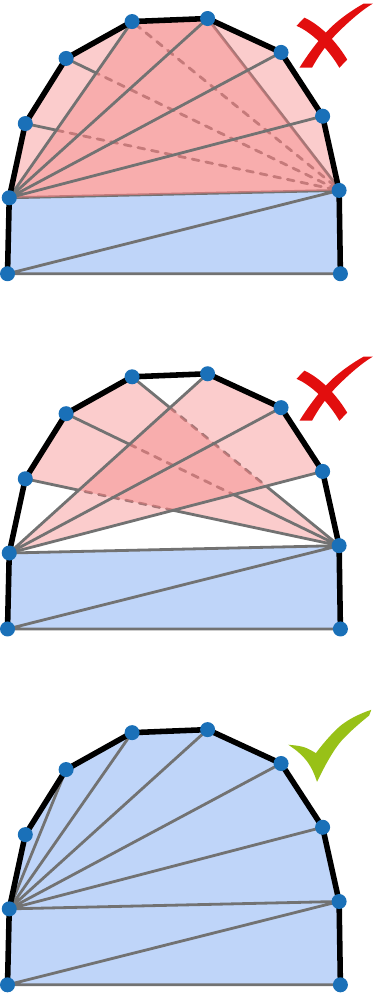}
  \end{center}
  \vspace{-10pt}
\end{wrapfigure}
Since strips often overlap along only a small portion of their boundaries, leaving one strip in place while deleting others would introduce unnecessary holes into the mesh. At the same time, deleting individual triangles next to non-manifold edges and vertices can result in an inconsistent mesh, which does not satisfy our persistence prior and contain undesirable holes and tunnels, as illustrated in the inset, middle. We obtain a manifold and fair solution that respects our priors by employing a correlation clustering framework~\cite{bansal2004correlation} (see inset,bottom). 
Our persistence term, combined with the use of the restriction of the matching set, strongly discourages the type of matches that lead to non-manifold artifacts.
Consequently, the non-manifold artifacts we face are typically very localized, allowing us to employ correlation clustering locally, one problematic mesh region at a time. 

We first identify pairs of  adjacent triangles which we consider as incompatible, namely ones that cannot jointly belong to the output mesh, using the following criteria (see inset).
 (1)  A  pair of triangles that share the edge $p_i,p_{i+1}$, are classified as incompatible if their non-shared vertices $q$ and $q'$ are on the same side of this edge.
 \setlength\columnsep{1pt}
\begin{wrapfigure}[11]{l}[0pt]{0.23\linewidth}
  \vspace{-15pt}
  \begin{center}
    \includegraphics[width=\linewidth]{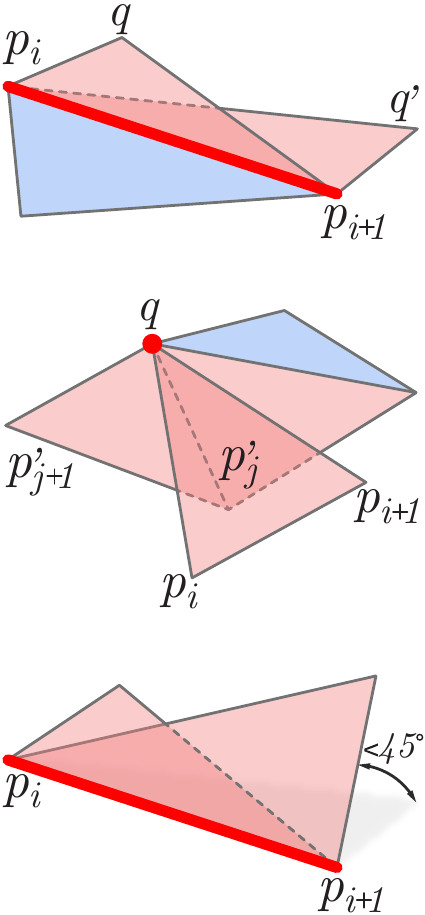}
  \end{center}
  \vspace{-10pt}
\end{wrapfigure}
(2) A pair of triangles \revised{$t_1=(q, p_i,p_{i+1})$ and $t_2=(q, p'_j,p'_{j+1}$) sharing} \old{that share} a common vertex $q$ \old{and two different edges $p_i,p_{i+1}$, $p'_j,p'_{j+1}$} are classified as incompatible
if they are on the same side of the stroke containing $q$ and if \old{either} the projection of one of the edges of \revised{$t_1$} \old{the first triangle $(q,p_i)$ or $(q,p_j)$} on the plane of \revised{$t_2$} \old{the second triangle $q,p'_j,p'_{j+1}$} intersects \revised{$t_2$} \old{this triangle} or vice versa (note that a valid mesh can contain multiple triangles that are on the same stroke side with respect to a common vertex as long as they do not ``overlap'').
(3) Lastly, while sharp creases in our output mesh are possible, we view them as undesirable, and classify triangles that share a common edge as incompatible if the dihedral angle between them is less than $45^{\circ}$. 

We resolve all of these artifacts by discarding a subset of the incompatible elements together with a subset of the triangles in their immediate vicinity producing a manifold mesh. In making the decision which elements to keep and which to remove we seek to maximize the output matching score $M$ (Equation~\ref{eq:global_score}). Since directly optimizing this score would make the problem intractable, we approximate it in our graph arc weight assignment as described below.

We compute the graphs we apply the clustering to as follows. 
We classify triangles as {\em undecided} if they belong to a set of incompatible triangles  or if they are immediately adjacent to an edge or vertex shared by a pair of incompatible triangles, and classify them as {\em output} otherwise. 
We form a separate graph for each connected component of undecided triangles (which share edges or vertices). Each graph has a node for each undecided 
triangle, and a single {\em output} node that represents all output triangles. We connect these nodes with arcs and assign arc weights as follows. 

--  We construct an arc for each pair of mutually incompatible triangles $t_1,t_2$, and assign it a high negative weight ($-30$).

--  We construct arcs for all pairs of undecided triangles that share common edges and are not mutually incompatible and assign them a weight of $1$.

-- We construct an arc between each undecided triangle and the output node and define its weight as follows. We recall that each triangle $t$ in our mesh connects a stroke edge $p_i,p_{i+1}$ to a vertex $q$ on the same or other stroke, which is to the left or right of this edge. We compute the sum of the matching scores $M(t) = M_a(p_i,q)+M_a(p_{i+1},q)$  where $a$ is $l$ or $r$ based on the side of the stroke that the vertex is on. We define the arc weight as $M(t) + C$ where $C$ is the number of edges shared by the triangle $t$ and output triangles. We include the edge count in the cost to implicitly minimize the size of the holes formed by the consolidation step. 

We use these \old{assignment} \revised{assignments} to formulate clustering as a constrained maximization problem. We maximize $\sum_{ij} w_{ij} Y_{ij}$, where $w_{ij}$ are the weights defined on the graph arcs, and $Y_{ij} = 1$ if the end nodes of an arc are in the same cluster and $Y_{ij} = 0$ otherwise. The sum  increases whenever the end-nodes of an arc with a positive weight are assigned to the same cluster or when end-nodes of an arc with a negative weight are kept apart.
We compute an adequate approximate solution to this problem using the lifted multicut approximation method~\cite{keuper2015efficient}. Following the computation, we retain the subset of undecided triangles that belong to the same cluster as the output node (Figure~\ref{fig:filtering}b). \revised{This subset is guaranteed to be manifold, as the correlation clustering method ensures that any pair of conflicting triangles are placed into different clusters.} \revised{We apply this clustering process to every group of triangles that are marked as undecided; the union of the triangles previously marked as output and the collection of subsets of undecided triangles kept after each clustering operation then form our output manifold mesh.}

\subsection{Partial Mesh Extension}
\label{sec:local_repeat}

\old{Our initial inter-stroke match computation restricts the matching candidate sets of the vertices along each stroke to reduce outlier matches (Section~\ref{sec:match}).}
\revised{The partial mesh generated via the three step process described above was computed by only considering matches from each stroke to itself and its dominant neighboring strokes.}
  This restriction produces mesh strips that satisfy all our criteria, but may leave side-by-side stroke sections unmatched in cases where a stroke has multiple immediately adjacent strokes on its left or right (Figure~\ref{fig:inter_stroke}b). We  connect such left-out stroke sections with mesh strips using a similar process to the one above. We first apply our matching algorithm (Section~\ref{sec:asses}) to  sections of the input strokes that lie on the boundaries of the current partial meshes. During the match computation, we restrict the candidate set of each vertex $p$ to include vertices on the boundaries of the connected mesh component that $p$ is on that satisfy \revised{the baseline matching conditions described} \old{(1) to (3)} in Section~\ref{sec:matching}, and use the same restriction on tangent similarity near end-vertices. The restriction to the same connected component is designed to limit the matches to lie on roughly similarly directed strokes. We then apply our consolidation process to the mesh computed from these matches. 
 Following \revised{this} consolidation, we have a manifold mesh (Figure~\ref{fig:inter_stroke}d), which connects similarly directed strokes using mesh strips. We compute consistent normal orientations for each connected component of this mesh using simple breadth-first-traversal and close obvious small holes (ones with four or less sides) inside each such component (typically located at transitions between different strips along the same stroke).


\section{Closing the Gaps}
\label{sec:global}

The final stage of our algorithm closes gaps between close-by components of the partial mesh as well as any remaining narrow holes within them. It achieves this goal by using a similar mesh strip formation process to the one used to form inter-stroke strips, with some minor differences outlined below. 
\revised{This step is quite similar to the mesh extension process (Section~\ref{sec:local_repeat}); while separating the two improves input fidelity, for simplicity of implementation the process outlined in Section~\ref{sec:local_repeat} can be skipped with only minor impact on fidelity.}

\paragraph{Boundary Smoothing.} The boundaries of the partial surfaces are often very jaggy and contain occasional overlaps between opposite boundaries. We resolve both artifacts by locally smoothing the boundary vertices, using the following simple update $p_i = p_i/2 + (p_{i-1} + p_{i+1})/4$. We only apply this update if it does not change significantly the normals of the adjacent triangles \revised{(which we evaluate by thresholding the angle between the pre-smoothed and smoothed normals to be at most $45^\circ$)}.  This step leads to more reliable matches and better shaped gap-spanning mesh strips. 

\paragraph{Matching.} 
We compute the matching scores as described in Section ~\ref{sec:asses}, defining the Frenet frame at each boundary vertex using the tangent to the boundary and the normal to the partial surface. We set the maximal distance $d_{max}$  for the boundaries of each connected partial surface component using the average of the distances between matched vertex pairs across this component. We define the candidate set of each vertex to include other vertices on the boundaries of \old{the}\revised{both this} partial surface \revised{and others} that satisfy \revised{ the baseline matching } conditions (1) and (3) in  Section~\ref{sec:matching}. We relax condition (2) to require the angle between $q-p$ and the boundary binormal $b(p)$ to be at most $80^\circ$ (we orient the binormal to point away from the bounded surface). We then proceed to compute the best matches for each boundary loop as described in Section~\ref{sec:matching}.

\paragraph{Meshing and Consolidation.} We form gap-spanning mesh strips by applying the algorithm in Section ~\ref{sec:meshing} as-is to the newly computed matches and remove non-manifold artifacts as described in Section~\ref{sec:global_filter}. During consolidation, we leave all triangles on the previously computed partial surface in place by labeling them as {\em output}.

\paragraph{Orientation} Our partial surfaces are oriented  during construction and we orient each gap-spanning strip after consolidation. 
However, when connecting these surfaces and strips together, we may introduce  gap-spanning strips that cannot be consistently oriented when merged with the connected components they bound (a Moebius strip effect).
We detect and resolve such configurations by first comparing the orientations of each partial surface and a strip it shared a border with.
For an orientable surface, we expect the orientations of pairs of border triangles (one from the strip and one from the partial  to either be identical for all pairs of triangles or be inverted for all of them. If this is not the case, we count the number of aligned and inverted pairs. We keep the strip triangles which conform with the majority choice (aligned or not) and discard the others. Finally, we use a greedy breadth-first traversal to establish a common orientation for the combined mesh. This process produces consistently oriented surfaces for all orientable input geometries tested.

\paragraph{Optional Post-Processing}
 Our core surfacing framework robustly closes narrow gaps and holes between input strokes. Since our system is not limited to closed surfaces only, the determination whether to surface larger and hence inherently ambiguous holes is left to the user.  We allow users to selectively close such large holes using  the hole-filling mechanism implemented in CGAL \shortcite{cgal:eb-18a}.  
Finally, users can smooth the resulting mesh using standard Laplacian smoothing to eliminate local geometric noise, and use Boolean operations to join intersecting closed mesh components together.

\revised{ As noted earlier, artists often use sparse strokes to communicate narrow geometries (such as the chicken's feet in Figure~\ref{fig:challenges}). We represent such isolated strokes, ones which are not part of the output triangulation, using their original triangulated ribbons.}

\section{VR Drawing Study}
\label{sec:user_study}
\label{sec:study}

\begin{figure}
 \includegraphics[width=\columnwidth]{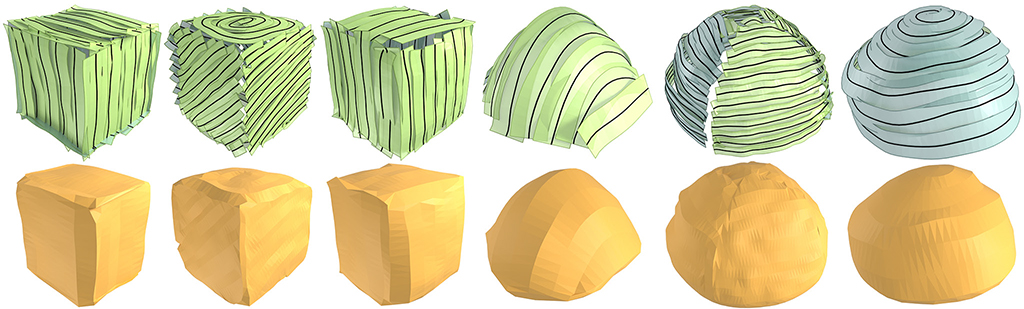} 
\caption{Typical drawings created by study participants when instructed to create cubes and half-spheres (domes) and our reconstruction results.}
\vspace{-15pt}
\label{fig:study}
\end{figure}

To observe how experts and amateurs communicate shapes when presented with a Virtual Reality flat stroke drawing interfaces, we asked five participants to draw simple shapes (cubes and half-spheres) using this interface. Our set of participants included one formally trained \old{artists} \revised{artist}, two 3D modelers, and two programmers. Each participant was provided with a quick tutorial on the use of the TiltBrush drawing interface; to avoid biasing the examples shown during the tutorial did not include any dense stroke drawings, but focused on basic TiltBrush manipulation. Participants were allocated a total of one hour to practice using TiltBrush and to then ``Try to draw a clean description of the surface'' of a cube and a half-sphere. They were told to stop once they were happy with the results. Three participants employed the dense side-by-side stroke  drawing style from the get-go and proceeded to draw both examples using this style. The remaining two created sparse curve drawings as their initial attempt, but were not satisfied with those and after some experimentation converged to the dense side-by-side stroke style as well.
Figure~\ref{fig:study} shows six of the drawings created by the participants. Additional drawings are included in the supplementary material.

\begin{figure*}
 \includegraphics[width=\textwidth]{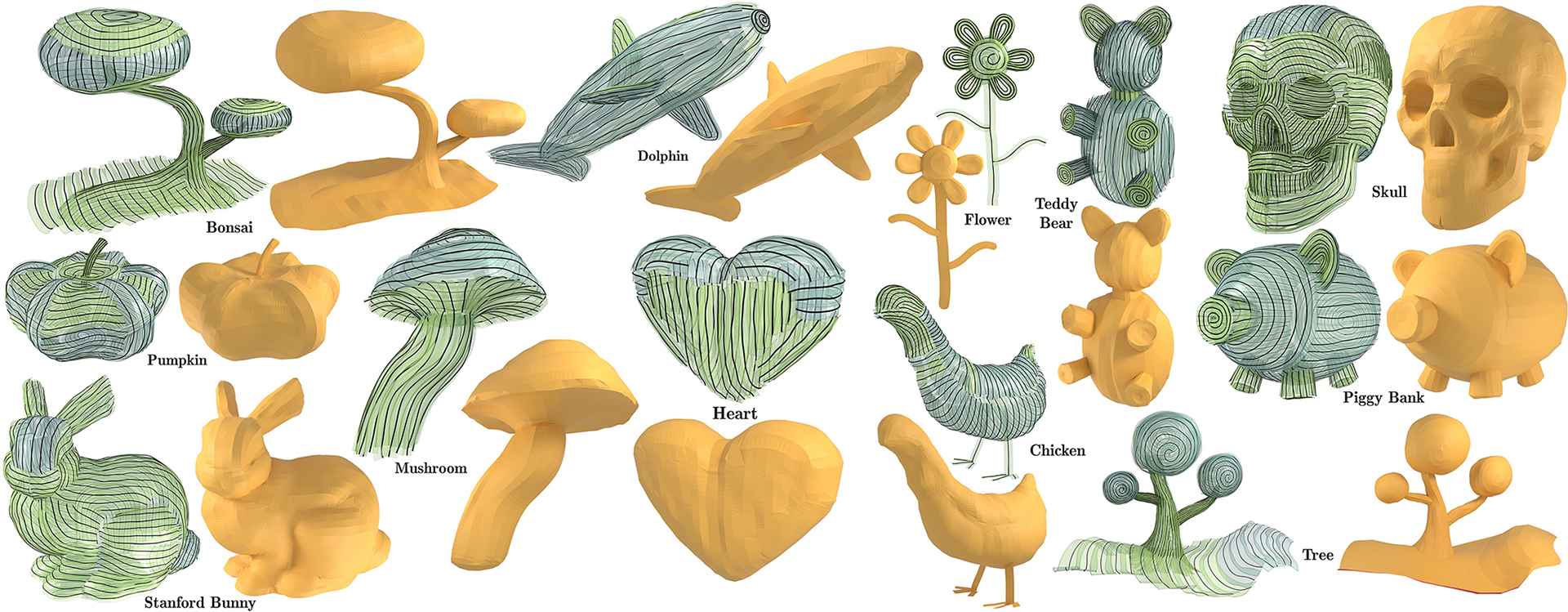} 
\caption{ A range of inputs and SurfaceBrush results. Flower and skull: $\copyright$ Enrique Rosales, bonsai, pumpkin, mushroom, dolphin, heart, chicken, teddy bear, piggy bank, tree: $\copyright$ Elinor Palomares.}
\label{fig:results}
\end{figure*}

In addition to this targeted experiment, we asked a modeler and a non-expert to create VR drawings of shapes of their choice using TiltBrush after showing them a few typical inputs created by one of the authors. The drawings they produced are shown in Figures~\ref{fig:teaser},~\ref{fig:compare}, and ~\ref{fig:results} (bonsai, pumpkin, mushroom, hart, tree, piggy bank, teddy, dolphin). 
The  drawings created exhibit the characteristics we describe in Section~\ref{sec:data}.
Drawing these models took the participants 30 minutes on average. Creating such irregular, free-form shapes using existing modeling technologies would require significantly more time and a degree of familiarity with these tools that our target users may not \old{posses} \revised{possess}. The professional  modeler who created our bunny input (Figure~\ref{fig:results}) in 25 minutes estimated that it would take him two and a half hours to create the same model in 3D Studio Max.
These experiments confirm that users see the dense side-by-side drawing style as a convenient and effective way to communicate shapes, and  validate our argument for employing dense stroke drawing as a modeling tool suitable for artists and amateurs.

\section{Results and Validation}
\label{sec:results}

We tested our algorithm on  \old{twenty seven} \revised{twenty nine} inputs. These include inputs created by amateur first-time users (Figure~\ref{fig:study}),  \old{a} \revised{an} amateur user who had some experience with the system (e.g. turtle, bonsai, chicken, heart), and two modelers (e.g. bunny, wooden horse, teapot, skull).  \revised{Two inputs (ship, plane) were downloaded from online repositories.} The inputs range in complexity from simple shapes (spheres and cubes in Figure~\ref{fig:study}) to complex models such as the bunny, horse, and skull. In all cases our outputs accurately reflect the user-drawn shapes. 

\setlength{\belowcaptionskip}{-5pt}
\setlength{\belowcaptionskip}{0pt}

\begin{figure}
\includegraphics[width=\columnwidth]{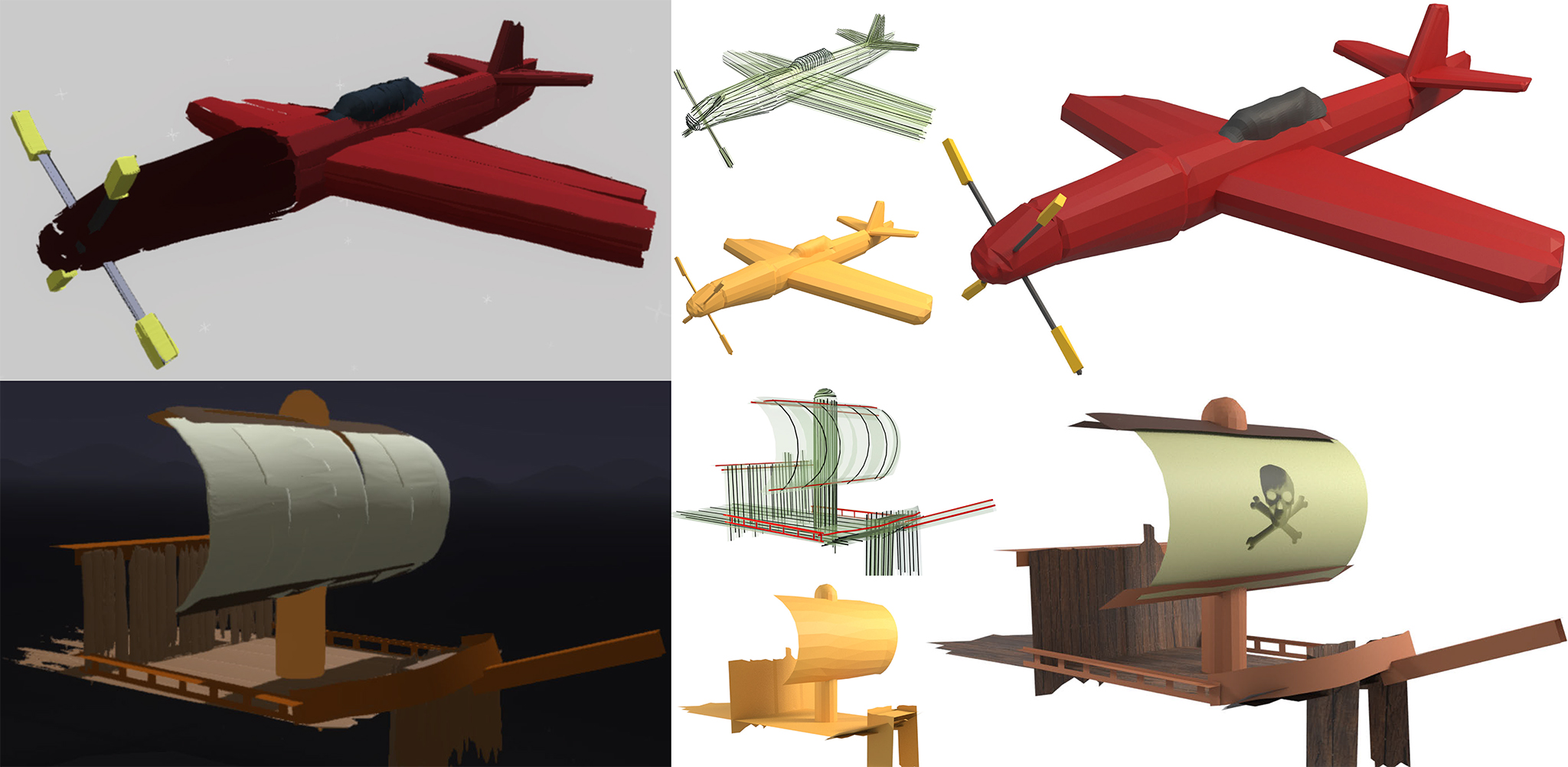} 
\caption{
\old{Given a coarse VR drawing from an online repository ($\copyright$ Skeazy J (poly.google.com)) we successfully reconstruct all the surfaces in the input while preserving isolated strokes (inset, red) intact. We use the obtained surface to rerender the input with texture (a functionality not supported by a ribbon-based representation).}
\revised{Given two VR drawing from an online repository ((top) $\copyright$ \textit{Olga Zinchenko} derived from \textit{I Have 2 Cents} (poly.google.com), (bottom) $\copyright$ Skeazy J (poly.google.com)) we successfully reconstruct all the surfaces in the input while preserving isolated strokes (inset, red) intact. We use the obtained surface (bottom, right) to rerender the input with texture (a functionality not supported by a ribbon-based representation).}}
\label{fig:online}
\end{figure}

\revised{
\paragraph{Optional Features.}}
Many images in online VR drawing repositories, see e.g. Figure~\ref{fig:online} are created to provide a compelling visual rather than a detailed model description; they contain multi-color strokes and use large numbers of isolated strokes to convey narrow ruled surfaces. 
 To process this data, we  augment our surfacing method to use color as a negative matching cue, disallowing matches between differently colored strokes. Our method reconstructs all surface elements, e.g. tower, sail, deck, in \old{this input} \revised{these inputs} while preserving the isolated features intact, allowing us to rerender  the \old{input} \revised{sail on the boat} with surface texture.

\begin{figure}
\includegraphics[width=\columnwidth]{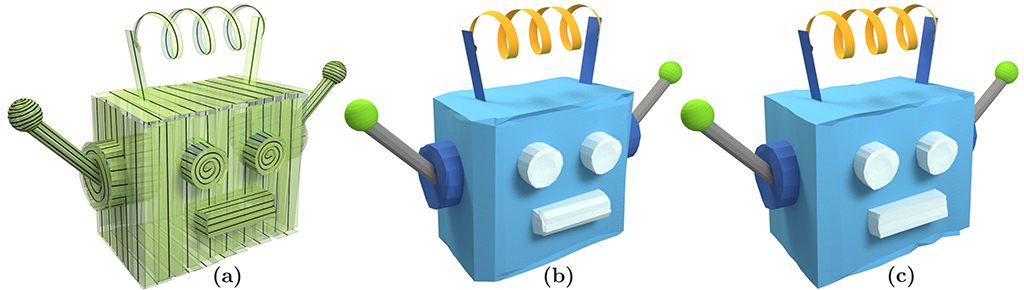} 
\caption{\revised{Optional sharp feature preservation: (a) input, (b) default output with sharp features rounded,  (c) with an optional feature preservation step. Input drawing: $\copyright$ Enrique Rosales.}}
\label{fig:corners}
\end{figure}

\revised{
Artists often depict surface creases (Figure~\ref{fig:corners}a) by drawing ribbons whose sides delineate the desired crease shape. Our default algorithm is designed to connect stroke spines and when used as is rounds such creases leaving a beveled edge (Figure~\ref{fig:corners}b). We provide users with an option to retain creases, if they choose to do so. To this end, when the the normals of two matched side-by-side strokes form angle of 90$^\circ$ or less, instead of connecting the stroke spines, we retain one half of the ribbon along  one of the strokes, and connect the side of that ribbon to the spine of the other. We use the stroke with a larger number of vertices along the 
marched section of interest to perform this task (Figure~\ref{fig:corners}c).}

\setlength{\columnsep}{1pt}
\begin{wrapfigure}{l}{0.2\linewidth}
\vspace{-15pt}
\centering
    \includegraphics[width=\linewidth]{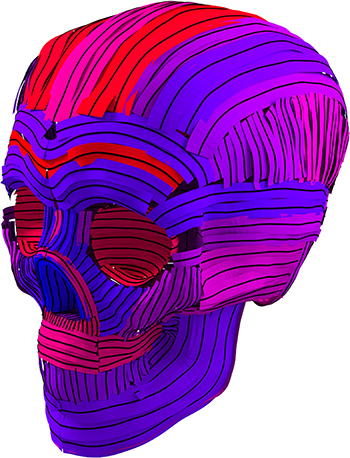}
\vspace{-25pt}
\end{wrapfigure}
\paragraph{Time Information.} VR drawing software records the time each stroke was drawn at. While \old{artist} \revised{artists} draw some immediately adjacent strokes sequentially, as the inset  shows, drawing order is not a reliable indicator of adjacency (color represents drawing time from earliest, blue, to latest, red). Thus limiting or biasing matches toward immediately preceding or succeeding strokes using drawing order, could produce undesirable artifacts on typical user inputs. Our framework is by design not dependent on stroke drawing order, and thus can robustly handle such typical data.

\begin{figure}
\includegraphics[width=\columnwidth]{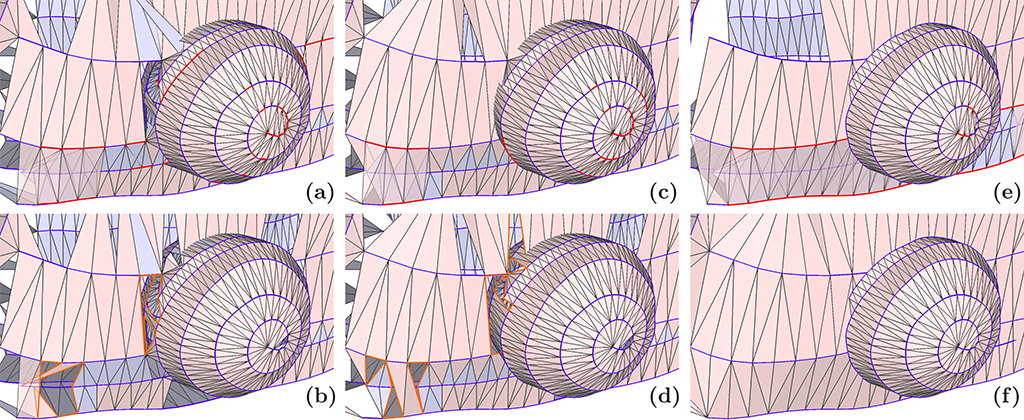} 
\caption{
Relying purely on \old{vertex to vertex} \revised{vertex-to-vertex} scores when computing matches between stroke vertices produces pre-consolidation partial meshes  (a) with numerous noisy matches and non-manifold edges resulting in post-consolidation surface with many artifacts (b).
Using unrestricted matches (c, before consolidation, d, after) produces less artifacts but leaves many undesirable holes, surfacing which would result in sharp dihedral angles and undesirable connections between separate mesh components. Our results before consolidation (e) and after partial surface computation (f). Our output is fair and keeps the separate components apart. Non-manifold edges in red, shading reflects front/back orientation.} 
\label{fig:pers}
\label{fig:without_with}
\label{fig:restricted_not}
\end{figure}

\paragraph{Comparison to Algorithmic Alternatives.} Figure~\ref{fig:pers} compares our method against two potential algorithmic alternatives. We show the impact of accounting for persistence during match assessment by introducing the persistence scores $S_e$ (Section~\ref{sec:asses})  by comparing our results (Figure ~\ref{fig:pers}f) to results produced using only vertex scores $S^v$ (Figure ~\ref{fig:pers}ab). As the figure shows, using vertex scores alone results in poor match persistence and subsequent visible surfacing artifacts. Similar artifacts appear if we do not restrict the first matching step (Section~\ref{sec:match}) to only dominant neighbor strokes (Figure~\ref{fig:without_with}cd). In all cases, the results obtained using our complete pipeline are significantly more reflective of the user input. 

\begin{figure*}
\includegraphics[width=\linewidth]{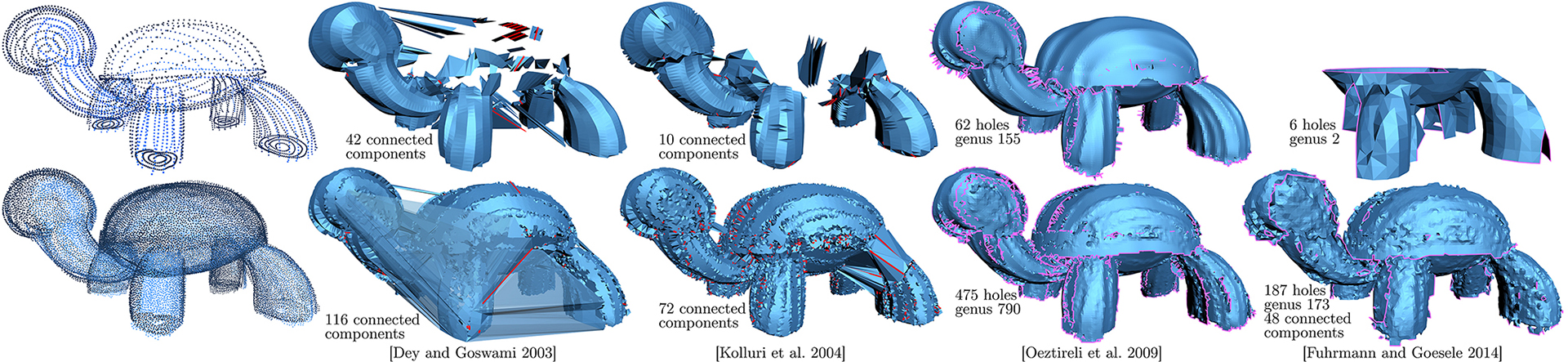} 
\caption{Comparison against representative point cloud reconstruction techniques using as input stroke vertices (top) and dense ribbon samples (bottom): Left to right : input, ~\cite{DeyG03}, \cite{Kolluri:2004},\cite{Oztireli09}, \cite{Fuhrmann}. Non-manifold edges highlighted in red, boundaries in purple. Our result shown in Figure~\ref{fig:turtle}. }
\label{fig:compare2}
\end{figure*}

\begin{figure*}
  \centering
  \includegraphics[width=\textwidth]{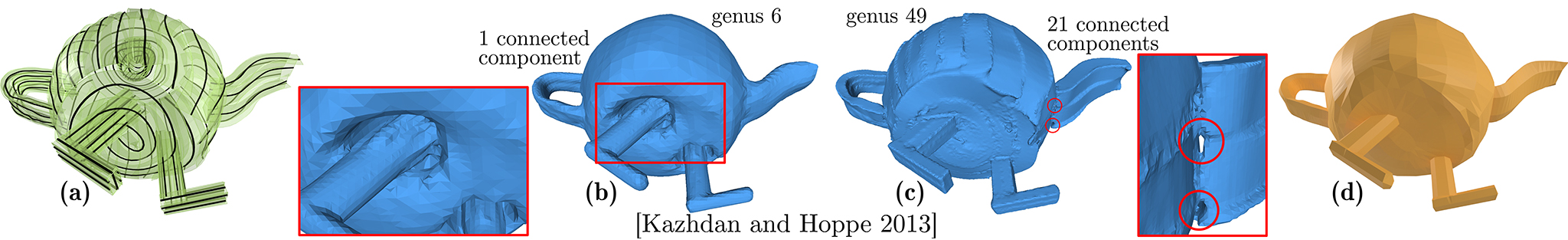} 
  \caption{Even with manually corrected normal orientation (a), state of the art frameworks that utilize normals, here~\cite{Kazhdan:2013}, produce outputs with excessive genus and other artifacts using stroke vertices (b) or dense ribbon samples (c). (d) Our result. Input drawing: $\copyright$ Jafet Rodriguez.}
  \label{fig:FixedNormals}
\end{figure*}

\begin{figure*}
\includegraphics[width=\linewidth]{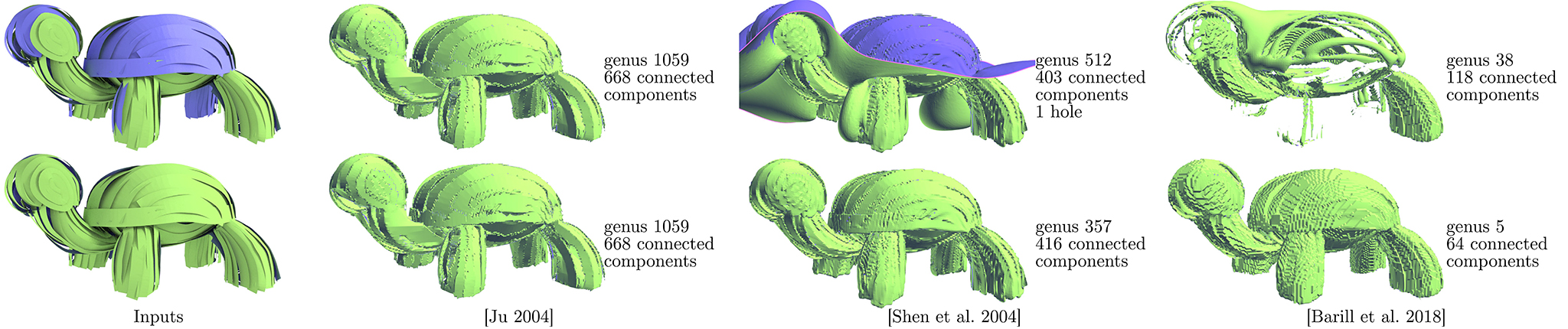} 
\caption{\revised{Attempting to reconstruct the user created models by applying topology repair methods to input ribbons fails on both original ribbons (top) and ribbons wit manually corrected normal orientation (bottom). Left to right: input, result of ~\cite{Shen:2004}, result of \cite{Ju:2004}, result of \cite{Barill:2018}. Our result is shown in Figure~\ref{fig:turtle}}. Input drawing: $\copyright$ Elinor Palomares.}
\label{fig:compare3}
\end{figure*}

\paragraph{Comparisons to Prior Art.}

As discussed in Section~\ref{sec:related} the inputs we process do not conform to the input specifications of the existing curve loop, cycle or network surfacing methods. Our input drawings can be easily converted into oriented point-clouds by using stroke polyline vertices or  point samples  on the ruled ribbons around them. Figures~\ref{fig:compare} and~\ref{fig:compare2} show comparisons of our outputs to those produced from such point-clouds using a range of state of the art techniques ~\cite{Kazhdan:2013,Oztireli09,Fuhrmann,Edels,DeyG03,Kolluri:2004}. Additional comparisons to reconstructions produced using these methods, \revised{and the ball-pivoting method of \cite{Bernardini:1999}} are included in the supplementary material. \revised{Ball-pivoting outputs exhibit similar artifacts to those demonstrated for the other reconstruction techniques.}

Existing methods that incorporate per-point normals as part of the data, e.g.~\cite{Kazhdan:2013,Oztireli09,Fuhrmann}, typically rely on those to have consistent in-out orientation. Our data does not satisfy this assumption, leading such methods to catastrophically fail, producing meshes with excessive genus and other artifacts. Even when the strokes are manually oriented for global consistency, the self-intersections between different stroke groups commonly present in our data cause major artifacts in the reconstructions computed with these methods (Figure~\ref{fig:FixedNormals}).
Delaunay type methods~\cite{Edels,DeyG03,Kolluri:2004} are similarly ill-suited for the uneven input spacing and the \old{low frequency} \revised{low-frequency} errors present in artist data, and produce outputs with large numbers of non-manifold edges and vertices, multiple redundant connected components, and with mesh triangles connecting unrelated surface parts. Our targeted framework overcomes all of these artifacts and produces the user-expected output on the tested inputs.

\revised{Topology Repair methods~\cite{Ju:2004,Shen:2004,Barill:2018} are designed for closed surfaces. Even when the input drawings depict closed shapes, applying these methods to our input triangle ribbons  produces inadequate results (Figure~\ref{fig:compare}, top). Result quality only marginally improves when ribbon orientation is  manually corrected  (Figure~\ref{fig:compare3}, bottom).}

\paragraph{Parameters and Runtimes} Our method has no user-tuned parameters, and all the results shown in the paper were obtained under identical conditions. Our algorithm takes under 5 seconds to surface the teapot (95 strokes, 3K vertices) \old{and similar complexity models} \revised{models of similar complexity}, and takes 52 seconds to surface our biggest model, the horse (298 strokes, 20K vertices)  on \old{an} a 4 core
Intel Core i7-6700HQ with  2.60\,GHz RAM and  32\,GB DDR4.  Out of this time, 60\% is spent in the matching code and 25\% doing consolidation.

\begin{figure}
\includegraphics[width=\columnwidth]{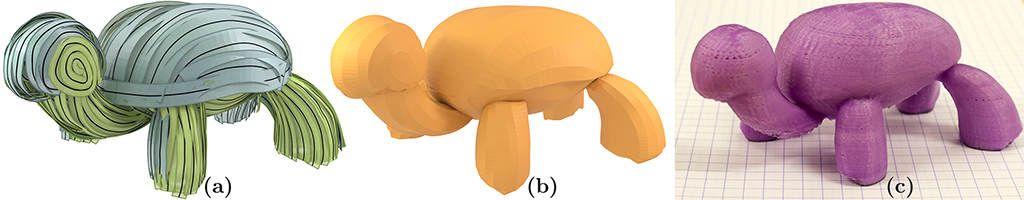} 
\caption{From VR drawing to 3D printed model: (a) input, (b) reconstructed surface, (c) 3D printed model. Input drawing: $\copyright$ Elinor Palomares.}
\label{fig:turtle}
\end{figure}

\paragraph{3D-Printable Models.}
As indicated earlier, and illustrated in Figures~\ref{fig:challenges} and~\ref{fig:components} users often use disjoint sets of strokes to draw different model parts; these are kept as disjoint components by our method. When a user specifies that the intended output is expected to be a connected closed intersection-free manifold surface, after closing all holes (Section~\ref{sec:global}) we use a Boolean union operation to combine all overlapping components into one. The resulting models can then be 3D printed as shown in Figures~\ref{fig:teaser},~\ref{fig:turtle}.

\begin{figure}
\includegraphics[width=\columnwidth]{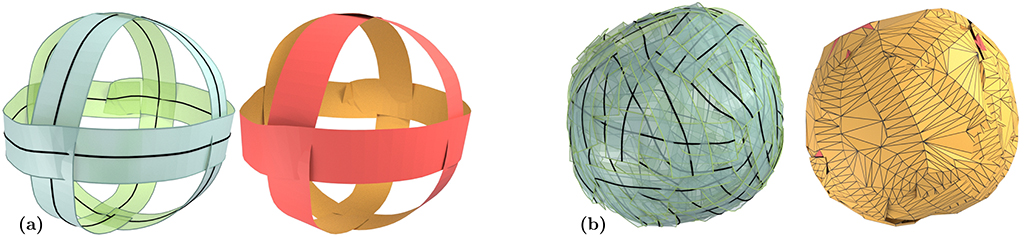} 
\caption{Our framework is not designed for surfacing very sparse (a) or randomly oriented (b) strokes.  In the former scenario it leave the input strokes essentially as is. In the latter where persistence and tangent consistency do not apply it produces results comparable to those obtained via reconstruction from point clouds.}
\label{fig:limitations}
\end{figure}

\section{Conclusions}
\label{sec:concl}

We present SurfaceBrush, a novel framework for freeform surface modeling using virtual reality brush strokes as input. 
This modeling interface is supported by a specialized surfacing algorithm that converts raw artist strokes into a manifold, user-intended surface. Our studies show that both experts and amateurs can successfully use our framework to create compelling 3D shapes. 

\paragraph{Limitations and Future Work.}  Our surfacing method is based on observations of practices artists typically employ when using VR brushes to draw 3D shapes. Thus it, predictably, breaks down when artists drastically deviate from the fence-painting metaphor  and use either very sparse (Figure~\ref{fig:limitations}a) or arbitrarily directed (Figure~\ref{fig:limitations}b) strokes. However, as our experiments show, even first-time users typically quickly converge to producing the type of inputs we expect \revised{when asked to depict geometric shapes}, and thus are unlikely to experience this limitation.  \revised{At the same time these restrictions may cause difficulties with processing of legacy inputs created to visually and artistically convey rather than model 3D content. Such inputs may use strokes to create artistic effects, e.g. mimicking van Gogh's, impressionist or Pointillist drawing styles, and may use layers of differently directed strokes to depict fur, hair, or texture. Our method is not designed to recover shapes from such highly stylized data.}

\revised{The focus of our reconstruction method is on fidelity to user input. Exploring regularization and beautification of  input and output created from non-expert drawings is an interesting future research topic that could potentially lead to more robust methods that reconstruct user intended rather than directly depicted shapes.}

Our reconstruction process is currently offline, thus users can only see the resulting model after completing the drawing. It would be interesting to explore  a variation of our method that provides users with real-time feedback as they draw. Such a method can potentially save user time and provide helpful real-time suggestions. As already noted, artist strokes are often aligned with principal (typically minimal absolute) curvature directions - it would be worth exploring how this extra information can be used in geometry optimization and other surfacing tasks down the line.

\begin{acks}
We are deeply grateful to Nicholas Vining and Nico Schertler for help with paper editing and proofing, to Luciano Silver Burla for help with video creation, to Chrystiano Ara\'{u}jo for help with code integration, to Elinor Palomares for her artistic inputs, and Chenxi Liu and Edoardo Dominici for help with running comparison experiments. The authors were supported by NSERC, CONACYT and Google. 
\end{acks}

\bibliographystyle{ACM-Reference-Format}
\bibliography{bibliography}

\end{document}